\documentclass[12pt]{article}

\usepackage{scicite}
\usepackage{graphicx}

\usepackage{times}

\newenvironment{sciabstract}{%
\begin{quote} \bf}
{\end{quote}}

\newcounter{lastnote}

\title{Superconductivity from buckled-honeycomb-vacancy ordering}

\author
{Yanpeng Qi,${}^{1\dagger\ast}$ Tianping Ying,${}^{2\dagger\ast}$ Xianxin Wu,${}^{3}$ Zhuoya Dong,${}^{1}$ Masato Sasase,${}^{2}$\\ Qing Zhang,${}^{1}$ Weiyan Liu,${}^{1}$ Masaki Ichihara,${}^{2}$ Yanhang Ma,${}^{1}$ Jiangping Hu,${}^{4}$\\ Hideo Hosono,${}^{2\ast}$\\
\\
\normalsize{${}^{1}$School of Physical Science and Technology, ShanghaiTech University, Shanghai 201210, China}\\
\normalsize{${}^{2}$Materials Research Center for Element Strategy, }\\
\normalsize{Tokyo Institute of Technology, Yokohama 226-8503, Japan}\\
\normalsize{${}^{3}$Institut für Theoretische Physik und Astrophysik,}\\ \normalsize{Julius-Maximilians-Universität Würzburg, Würzburg 97074, Germany}\\
\normalsize{${}^{4}$Beijing National Laboratory for Condensed Matter Physics and Institute of Physics,}\\
\normalsize{Chinese Academy of Sciences, Beijing 100190, China}\\
\\
\normalsize{${}^{\dagger}$Y.Q. and T.Y. contribute equally to this work.}\\
\normalsize{$^\ast$E-mail: qiyp@shanghaitech.edu.cn; t-ying@mces.titech.ac.jp; hosono@mces.titech.ac.jp}
}

\date{\today}

\begin{document} 


\baselineskip24pt


\maketitle


\begin{sciabstract}
  Vacancies are prevalent and versatile in solid-state physics and materials science. The role of vacancies in strongly correlated materials, however, remains uncultivated until now. Here, we report the discovery of an unprecedented vacancy state forming an extended buckled-honeycomb-vacancy (BHV) ordering in Ir$_{16}$Sb$_{18}$. Superconductivity emerges by suppressing the BHV ordering through squeezing of extra Ir atoms into the vacancies or isovalent Rh substitution. The phase diagram on vacancy ordering reveals the superconductivity competes with the BHV ordering. Further theoretical calculations suggest that this ordering originates from a synergistic effect of the vacancy formation energy and Fermi surface nesting with a wave vector of (1/3, 1/3, 0). The buckled structure breaks the crystal inversion symmetry and can mostly suppress the density of states near the Fermi level. The peculiarities of BHV ordering highlight the importance of “correlated vacancies” and may serve as a paradigm for exploring other non-trivial excitations and quantum criticality.
\end{sciabstract}


\section*{Introduction}

Crystals inherently possess imperfections. Vacancies, as the simplest form of point defects, significantly alter the optical, thermal, and electrical properties of materials\cite{mccarty2001}\cite{pertsinidis2001}. Well-known examples include colour centres in many gemstones, the nitrogen-vacancy centre in diamond \cite{childress2006,mamin2013}, vacancy migration in solid-state batteries \cite{li2014,wang2018}, and the metal-insulator transition in phase-change materials \cite{salinga2011,loke2012}. The vacancies in these cases are in frameworks with no or weak interactions. However, the role of vacancies in strongly correlated materials is thus far unclear due to the lack of an ideal prototype.

Strongly correlated vacancy ordering has long been anticipated to harbour exotic physics, such as superconductivity. The K-Fe-Se superconductor has been a hot research subject in recent studies for an important reason, viz., the existence of an insulating iron-vacancy-ordering phase \cite{guo2010,dagotto2013}. However, this vacancy-ordering phase has been proven to coexist with the superconducting phase at the nanoscale but is not responsible for the superconductivity (as will be discussed later). Whether correlated vacancies could become a new type of superconducting parent phase is an unanswered question. Iridates, with comparable and competing energy scales of the on-site Coulomb repulsion, crystal field and spin-orbit coupling, are a platform of rich structures and physical properties \cite{radaelli2002,tsong1992,kim2009,kitagawa2018,qi2012}. A comprehensive investigation into the Ir-Sb binary system leads us to discover a new type of vacancy ordering that serves as the first superconducting parent phase with order vacancies.

\begin{figure*}[t]%
\includegraphics[width=16cm]{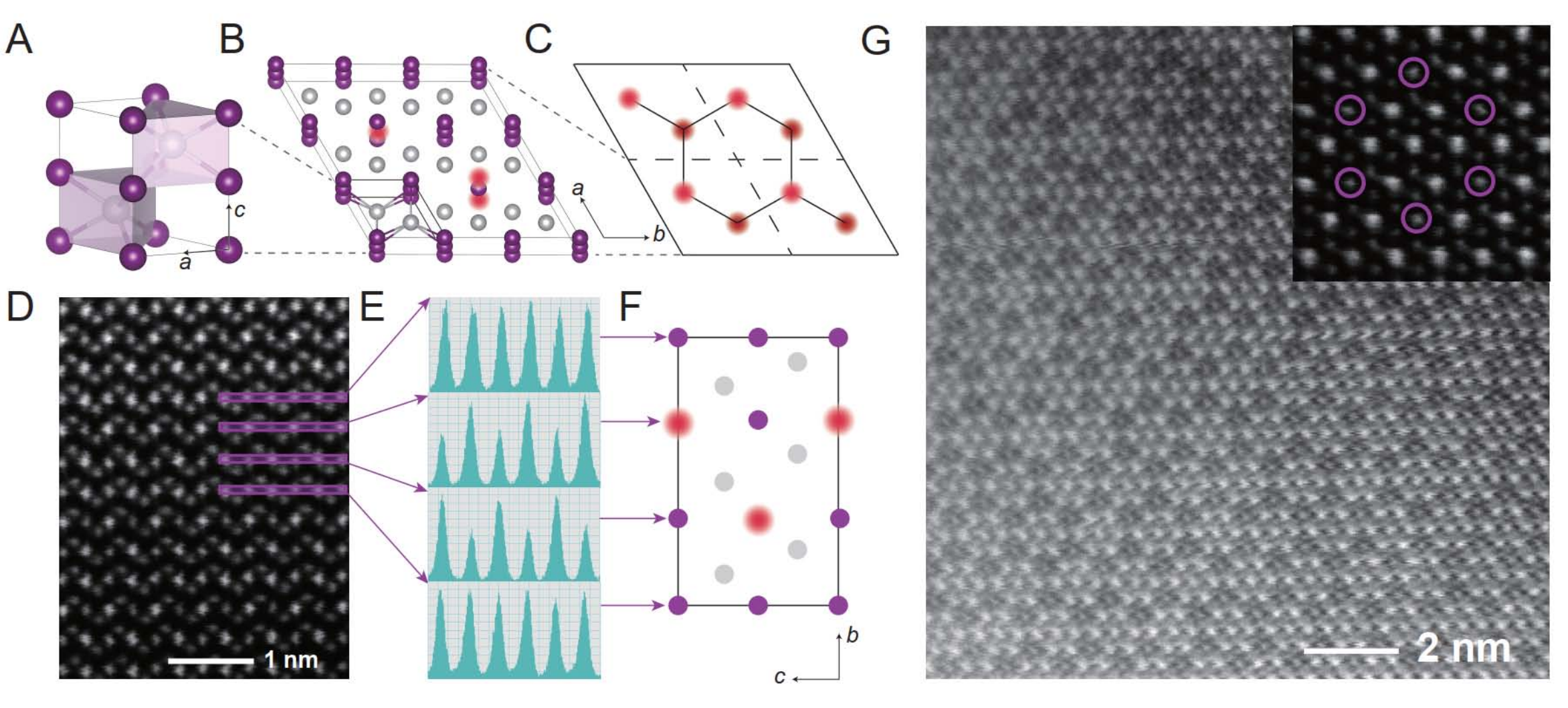}
\caption{\label{fig1}  \textbf{Superstructure with BHV ordering.} (\textbf{A}) NiAs-type primary cell (space group $P6_3/mmc$). The point groups at the Ir (purple circles) and Sb (grey circles) sites are $D_{3d}$ and $D_{3h}$, respectively. (\textbf{B}) Crystal structure of Ir$_{16}$Sb$_{18}$. Ir vacancies are indicated by red shadowed circles. (\textbf{C}) Illustration of BHV ordering along the $c$-axis. Note that the Ir and Sb atoms are omitted for clarity. (\textbf{D}) Aberration-corrected scanning transmission electron microscopy (STEM) micrograph along the $a$-axis. (\textbf{E}) Intensity line profile of the regions marked by arrows. (\textbf{F}) Projection of the BHV superstructure along the $a$-axis. (\textbf{G}) STEM micrograph of Ir$_{16}$Sb$_{18}$ along the $c$-axis. The dark sites are the iridium vacancies. The inset shows one honeycomb hexagon at a higher magnification.}
\end{figure*}

\section*{Results}

\begin{figure*}[htbp]%
\includegraphics[width=16cm]{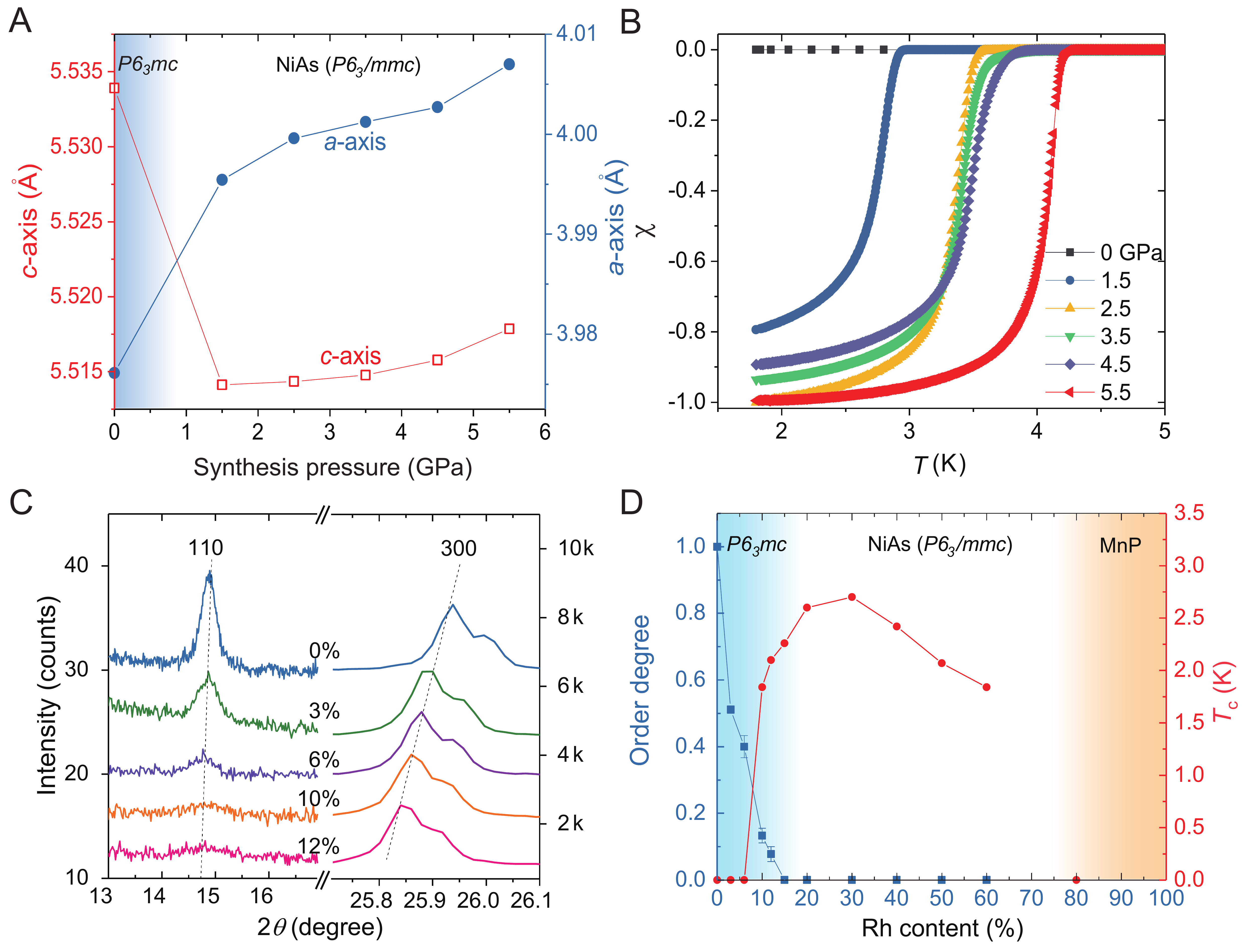}
\caption{\label{fig2} \textbf{Competition between BHV ordering and superconductivity.} (\textbf{A}) Lattice parameters along the $a$- and $c$-axes of Ir$_{1-\delta}$Sb with increasing synthesis pressure. The sample at 0 GPa was acquired by arc melting. The $a$-axis of Ir$_{16}$Sb$_{18}$ is normalized for comparison with that of the NiAs primary cell. (\textbf{B}) Diamagnetization (calibrated using Pb as a standard) of Ir$_{1-\delta}$Sb samples synthesized under various pressures. (\textbf{C}) Intensity of the Laue 110 (superstructure) and 300 (main structure) reflections at various isovalent Rh contents. The curves are vertically shifted for clarity. (\textbf{D}) Phase diagram of Ir$_{1-x}$Rh$_x$Sb. Here, the order degree is defined as the normalized peak ration of (110/300) with the reference to that of the Ir$_{16}$Sb$_{18}$ ($x$ = 0 \%). Superconductivity emerges with the collapse of the BHV ordering and peaks at approximately $x$ = 30 \%. Upon further increases in the Rh content, $T_c$ smoothly decreases until terminating abruptly at approximately $x$ = 80 \%, at which point the material undergoes a first-order transition from the NiAs- to MnP-type structure.}
\end{figure*}

\subsection*{Superstructure with BHV ordering}
The emergence of additional ordered states was initially discovered by x-ray diffraction (XRD), from which a series of systematic superstructure peaks can be distinguished. Through density functional theory (DFT) calculations, electron diffraction (ED), aberration corrected scanning transmission electron microscopy (STEM) and Rietveld refinement of the XRD pattern, the BHV-ordering phase is determined to be a high-temperature phase with a space group of $P6_3mc$ (Supplementary Materials \textbf{1-2}). This phase can be viewed as a supercell with 3$\times$3$\times$1 NiAs-type building blocks (Figs. 1A and 1B). The positions of Ir vacancies in the tripled supercell are revealed by the atomic-resolution image captured along the $a$-axis and the intensity line profiles (Figs. 1D-1F). Remarkably, a honeycomb arrangement can be seen in Fig. 1G, which is a projection of the buckled honeycomb pattern of Ir vacancies along the $c$-axis (Fig. 1C). To the best of our knowledge, the BHV ordering pattern discovered in Ir$_{16}$Sb$_{18}$ has no precedent in reported compounds.

The BHV ordering described here is distinct in its absence of an inversion centre compared to its NiAs building block and planar honeycomb structures. Usually, various magnetic (spin) and charge orderings in strongly correlated materials, such as cuprates and iron pnictides, can only break the lattice rotational symmetry and preserve the inversion symmetry. In contrast, the BHV ordering is one of the rare cases that exhibit the opposite trend, which may result in unexpected phenomena. A straightforward way to examine the effect of vacancy ordering on the physical properties is to squeeze extra Ir into the vacant sites under high pressure.

\subsection*{Competition between BHV ordering and superconductivity}

Rietveld refinement of the Ir$_{1-\delta}$Sb synthesized at the lowest pressure of 1.5 GPa (Supplementary Materials \textbf{3}) reveals that the BHV ordering is completely suppressed. The expansion seen along the $a$-axis is consistent with the expected filling of the vacancy sites (Fig. 2A). However, an abrupt drop in the low-pressure region can be observed along the $c$-axis. This drop arises because two adjacent layers with charged BHV ordering tend to repel each other, and the squeezing of extra Ir atoms into the vacant sites under pressure randomizes the BHV ordering, reduces the Coulomb repulsion force of the interlayers and collapses the lattice. At the same time, the $P6_3mc$ (Ir$_{16}$Sb$_{18}$) phase gradually transforms into $P6_3/mmc$ (Ir$_{1-\delta}$Sb). Superconductivity emerges following the collapse. As shown in Fig. 2B, a sharp diamagnetic onset transition at 2.8 K appears in the sample synthesized at 1.5 GPa and continuously increases to 4.2 K at 5.5 GPa. The Rietveld refinement of the superconducting samples are shown in Figs. S8 and S9. The zero resistivity and heat capacity are depicted in Figs. S10 and S11, respectively.

\begin{figure*}[ht]%
\includegraphics[width=16cm]{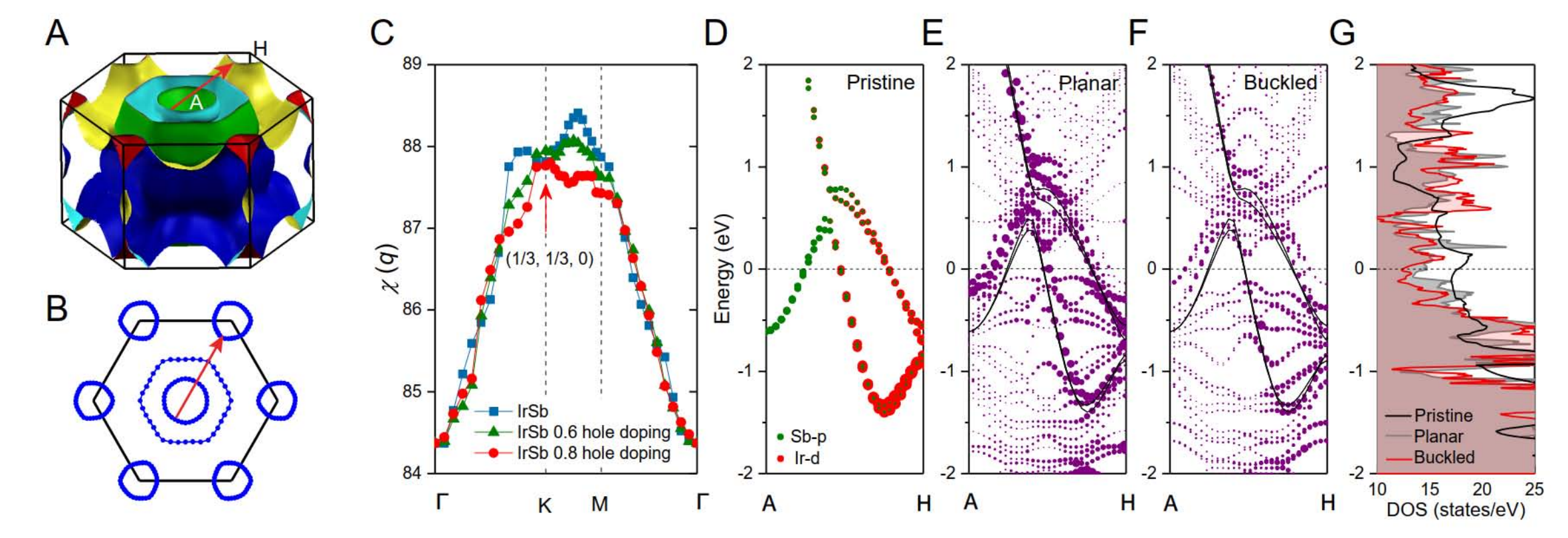}
\caption{\label{fig3} \textbf{Origin of the BHV ordering.} (\textbf{A}, \textbf{B}). Fermi surface topology and 2D cut in the $k_z$ = $\pi$ plane of pristine IrSb. The red arrows represent the nesting vector of $q$ = (1/3, 1/3, 0). (\textbf{C}) Charge susceptibility of Ir$_{1-\delta}$Sb with hole doping. The calculation is based on the pristine IrSb structure with a rigid-band shift of the Fermi level. The peaks around point $K$ are attributed to nesting in a and b and are further shifted towards point $K$ as the extent of hole doping increases. (\textbf{D}) Orbital-resolved band structure of pristine IrSb along the $A$-$H$ direction. The electron pocket around point $H$ is attributed to the $d$-orbital of Ir, while the other pocket around point $A$ is due to the hybridization of the $p$- and $d$-orbitals of Sb and Ir, respectively. (\textbf{E}, \textbf{F}). Band unfolding of the planar honeycomb and BHV superstructures in a NiAs-type unit cell. The sizes of the circles are proportional to the spectral weights. The bands of pristine IrSb are superimposed as solid black lines. (\textbf{G}) Density of state (DOS) of the pristine IrSb (black), planar honeycomb vacancy (grey), and BHV (red) configurations.}
\end{figure*}

The BHV ordering is completely suppressed at 1.5 GPa (Fig. S8); the most intriguing region where the BHV ordering tails off and superconductivity gradually emerges remains unexplored. A feasible tuning method is isovalent substitution with Rh at the Ir sites. A series of Ir$_{1-x}$Rh$_x$Sb samples, which have a solid solution of Rh up to 60 \%, can be easily acquired via arc melting at ambient pressure (Fig. S12). 

As shown in Fig. 2C, the 110 peak of the superstructure gradually decreases and vanishes at 12 \%, while the continuous shift of the 300 peak suggests random Rh substitution, agreeing well with the composition image (Fig. S13). A phase diagram is summarized in Fig. 2D, and a detailed characterization of Ir$_{1-x}$Rh$_x$Sb can be found in Supplementary Materials \textbf{4}. It is interestingly to observe a superconducting dome between the two non-superconducting end points. Although distinguishing the superstructure diffraction peaks beyond $x$ = 12 \% is difficult, the short-range order of the vacancy interactions should persist up to a critical point where strong bosonic fluctuations lead to the $T_c$ maximum. 

\subsection*{Origin of the BHV ordering}

To elucidate the possible origin of the BHV ordering in Ir$_{1-\delta}$Sb, we studied the corresponding electronic structures. The Fermi surfaces of pristine IrSb and a two-dimensional cut along the $k_z$ = $\pi$ plane are displayed in Figs. 3A and 3B. The charge susceptibility $\chi_0$ can be calculated by the following equation:
\begin{equation}
\chi_0(q) = \frac{1}{N}\sum_{pnm}\frac{{f(\epsilon_{np+q})-f(\epsilon_{mp})}}{\epsilon_{mp}-\epsilon_{np+q}}
\end{equation}
where $N$ is the number of $k$-points in the discretized Brillouin zone, $\epsilon_{np}$ is the energy of band $n$ at momentum $p$, and $f$ is the Fermi-Dirac distribution. In the $k_z$ = $\pi$ plane, we observed relatively good nesting between the inner circular pockets and hexagonal pockets around points $A$ and $H$, respectively, as shown in Fig. 3B. This nesting contributes to the peaks in the susceptibility around point $K$ (Fig. 3C). The nested pockets are of the electron type (Fig. 3D). This nesting is robust, and slight hole doping (0.6 and 0.8 holes) originating from the random Ir vacancies can further shift the peaks towards point $K$ (Fig. 3C). Upon further inclusion of interactions, the susceptibility will diverge at point $K$, resulting in a charge-density-wave (CDW) instability with a 3$\times$3$\times$1 supercell. Similar charge fluctuation-mediated superconductivity has also been observed in some organic superconductors \cite{ishiguro2012,merino2001}. 

Although the calculated nesting vector (1/3, 1/3, 0) agrees well with the experimentally observed 3$\times$3$\times$1 supercell, the low-energy electron interaction alone (within several hundred kelvin) cannot completely explain the observed BHV ordering quenched from high temperature. This energy discrepancy prompts us to further take the vacancy formation energy into consideration. Within a 3$\times$3$\times$1 framework, we consider different types of defects (Ir/Sb vacancies and Ir/Sb interstitial atoms), number of vacancies and their arrangements. According to Figs. S18 to S20, only the BHV ordering best maximizes the vacancy-vacancy distance in the given 3$\times$3$\times$1 framework, which consequently further decreases the total energy. The synergistic effect of the vacancy formation energy and Fermi surface nesting highlights the correlation among vacancies and provides a plausible origin for the experimentally observed BHV ordering in Ir$_{16}$Sb$_{18}$. Moreover, the unfolded band structure of the BHV ordering (Fig. 3F, for details, see methods) was found to be quite smeared compared to that of a similar structure consisting of a planar honeycomb arrangement (Fig. 3E). The suppressed spectral weights in the BHV ordering indicate that a large amount of the electron spectral weight near the Fermi surface is transferred (Fig. 3G). The full range of the unfolded bands can be found in Fig. S21, together with the other vacancy configurations.

The combination of a competing phase, Fermi surface nesting, and the dome-shaped phase diagram of $T_c$ highlights the uniqueness of the observed superconductivity. Notably, the characteristics of Ir$_{1-x}$Rh$_x$Sb are in sharp contrast to those of potassium-intercalated FeSe, in which nano-scale coexistence of the insulating vacancy-ordering phase K$_2$Fe$_4$Se$_5$ (245) and the iron fully occupied K$_x$Fe$_2$Se$_2$ superconducting phase were discovered. It is a phase separation in K-Fe-Se system without intermediate states \cite{li2012,shoemaker2012}. That’s why K-doped FeSe, as well as other intercalated FeSe-based superconductors, exhibits fixed $T_c$ \cite{guo2010} and lattice parameters \cite{ying2013}. A crucial criterion for a material to be called a superconducting parent phase is whether or not one can find a competing order such as the charge density wave in cuprates and spin density wave in iron pnictides. Various routes have been tried to tune continuously from the 245 phase to the superconducting phase but failed. Now it's a consensus that the insulating 245 phase coexists with the superconducting phase at the nanoscale but is not responsible for the superconductivity \cite{li2012,shoemaker2012,yuan2012wangnanlin,ying2013,burrard2013,sedlmaier2014ammonia,sun2019intercalating}. We also note the widely investigated Na$_x$CoO$_2$ showing superconductivity when hydrated \cite{takada2003,schaak2003,merino2006,roger2007}. Although often described as Na ordering, it could equally be viewed as vacancy ordering. Apart from the poorly tuneable water content, the crystal structures of the hydrate and anhydrate forms are very different. Thus, the discovered clear competition between the BHV ordering with the superconductivity shown in Fig. 2D renders it the first identified superconducting parent phase with correlated vacancy ordering.

\section*{Conclusion}
The discovery of BHV ordering is the most remarkable aspect of this material. The buckled structure effectively breaks the inversion symmetry of the crystal and minimizes the electrons near the Fermi surface. Our study suggests that the ordered vacancy can be a new degree of freedom for the manipulation and study of quantum materials. Investigating how the vacancy intertwines with other conventional degrees of freedom, such as lattice, spin and orbital, to affect the physical properties of the materials will be extremely interesting. For example, further studies are required to reveal how the buckled-vacancy structure affects the vibrational phonon modes and generates a possible non-trivial spin texture in the vacancy sites due to a geometrically frustrated vacancy structure. Investigations into the BHV-ordering phase and other structures with “correlated vacancies” may hold promise for novel discoveries in physics.

\section*{Materials and Methods}

High purity Ir, Rh, and Sb powders (Kojundo Chemical Laboratory Co. Ltd.) were weighed and mixed in a glove box filled with purified Ar gas (H$_2$O, O$_2$ $<$ 1 ppm). For all the arc-melting and high-pressure samples, the nominal ratio of Ir(Rh) : Sb = 1 : 1. A high-pressure synthesis process was conducted in a belt-type high-pressure apparatus \cite{belt-type}. The pressed pellets were placed in an $h$-BN capsule and heated at 2073 K under various pressures (1.5-5.5 GPa) for 2 h before being quenched to room temperature. Samples with a nominal composition of Ir$_{1-x}$Rh$_x$Sb were synthesized via the arc-melting method.

Powder x-ray diffraction (XRD) patterns were obtained using a Bruker D8 Advance diffractometer with Cu-$K_\alpha$ radiation (wavelength $K_{\alpha1}$ = 1.54056 \AA, $K_{\alpha2}$ = 1.544390 \AA) at room temperature. The FullProf software package \cite{fullprof} and commercial software Jade Pro 7.8.2 were used for the Rietveld refinement. Electron diffraction data were collected using a JEM-2100Plus microscope. High-angle annular dark-field (HAADF) and transmission electron microscopy (TEM) images were captured using a JEM-ARM200F and a GrandARM300F. Transport, magnetic, and thermodynamic properties were measured using a physical property measurement system (PPMS, Quantum Design) and a superconducting quantum interference device (SQUID) vibrating sample magnetometer (SVSM, Quantum Design). The elemental compositions were determined by probing approximately 50 focal points using an electron-probe microanalyser (JEOL, Inc., model JXA-8530F), the results of which were averaged.

In the density functional theory (DFT) calculations, we applied the projector augmented wave (PAW) method encoded in the Vienna Ab Initio Simulation Package (VASP) \cite{vasp1,vasp2,vasp3} to describe the wave functions near the core and the generalized gradient approximation (GGA) used in the Perdew-Burke-Ernzerhof (PBE) parameterization to determine the electron exchange-correlation functional \cite{pbe}. We relaxed the lattice constants and internal atomic positions with a plane wave cutoff energy of 450 eV, and the forces were minimized to less than 0.01 eV/\AA. In the vacancy formation energy calculation, 3$\times$3$\times$2 superstructures of the NiAs primary cell, e.g. Ir$_{18}$Sb$_{18}$ and Ir$_{17}$Sb$_{18}$, were used. The number of $k$-points in the Monkhorst-Pack scheme \cite{mpmash} was 4$\times$4$\times$4 for structure relaxation and 6$\times$6$\times$6 for self-consistent calculations. Note that spin-orbit coupling (SOC) was taken into account in the calculations. The entropy was accounted for using the open source software package $Phonopy$ \cite{phonopy}. Band unfolding of the buckled-honeycomb-vacancy (BHV) ordering, planar, honeycomb, and anti-diagonal configurations and calculations of their respective spectral weight were performed using the $vasp\_unfold$ software package \cite{vaspunfold}. When calculating the susceptibility, 32 maximum localized Wannier functions (with SOC) including Ir $d$-orbitals and Sb $p$-orbitals were used in the construction of the tight-binding model.

\noindent \textbf{Acknowledgements:} 
We thank Y.Muraba and Y.Kondo for experimental help and T.Kamiya, T.Inoshita, J.J.Lin, T.X.Yu, H.C.Lei, O.Terasaki, W.Zhang, and S.Haindl for fruitful discussions. We especially thank S.F.Jin from IOP(CAS), R.S.Zhou from International Centre for Diffraction Data-MDI and C.H.Xu from International Centre for Diffraction Data Beijing office for their help on the Rietveld refinement of the BHV ordering phase. \\
\noindent \textbf{Funding:}  Y.Q. acknowledges the support by the National Key R$\&$D Program of China (Grant No. 2018YFA0704300), the National Science Foundation of China (Grant No. U1932217 and 11974246) and Natural Science Foundation of Shanghai (Grant No. 19ZR1477300). This study was supported by MEXT Element Strategy Initiative to form Core Research Center and partially supported by C$\hbar$EM, SPST of ShanghaiTech University (Grant No. 02161943) and Analytical Instrumentation Center (contract no. SPST-AIC10112914), SPST of ShanghaiTech University. J.H. is supported by National Science Foundation of China (Grant No. NSFC-11888101).\\
\noindent \textbf{Author Contributions} Y.Q., T.Y. and H.H. conceived the project. T.Y., Y.Q. did majority of the sample preparation and measurements with the help of Y.M.. T.Y, and X.W. performed the DFT calculation. Z.D., Y.M., Q.Z., M.S., and M.I. carried out the STEM images. Y.K. did the EMPA measurements. T.Y., Y.Q., X.W., and H.H. analyzed the data and wrote the manuscript with the contribution of all the authors.\\
\noindent \textbf{Competing Interests} The authors declare that they have no competing financial interests.\\
\noindent \textbf{Data and materials availability:} Additional data and materials are available online.

\bibliography{scibib}

\pagebreak
\begin{center}
\textbf{\large Supplemental Materials of \\``Superconductivity from buckled-honeycomb-vacancy ordering"}
\end{center}
\section{High-temperature phases in the Ir-Sb binary system}

The hexagonal Ir-Sb binary compound was first reported in 1957 \cite{caillat1957}, the detailed phase diagram of which can be found in Ref. \cite{caillat1993}. In our initial work, we failed to obtain the hexagonal phase after heating the mixture even to above 2000 $^{\circ}$C in radio frequency (RF) or spark plasma sintering (SPS) heating systems. The remnants after heating consisted of Ir and IrSb$_2$ without any trace of the hexagonal phase. However, we successfully obtained the hexagonal phase using arc melting. Compared to RF heating and SPS, the major difference in arc melting is its fast cooling rate, which implies that the hexagonal phase may actually be a high-temperature phase. As the preferred crystal structure has the lowest Gibbs free energy, inclusion of the entropy contribution is necessary to gain a deeper insight into the reaction.

Standard DFT calculations are understood to be only valid at $T$ = 0 K, and the formation energy is defined as $\Delta\epsilon$ = $E_{hex}$ - $E_{Ir}$ - $E_{IrSb_2}$. We included the temperature-dependent vibrational free energy $F_{vib}$ in the calculations, and thus, the formation energy can be written as $\Delta G$ =$\Delta\epsilon$ + $\Delta F_{vib}$ = $E_{hex}$ - $E_{Ir}$ - $E_{IrSb_2}$ + $F_{hex}$ - $F_{Ir}$ - $F_{Irsb_2}$. “Hex” here represents Ir$_{16}$Sb$_{18}$ at ambient pressure or IrSb at 5 GPa. As shown in Fig. S1, the formation energy is positive in the low temperature region, which indicates that the hexagonal phases will automatically decompose into Ir and IrSb$_2$ when slowly cooled. In contrast, the hexagonal phases are stable at high temperature due to the entropy contribution and can be obtained by quenching. This was further confirmed by annealing the $P6_3mc$ phase at 800 $^{\circ}$C (in a sealed quartz tube) or slowly cooling the $P6_3/mmc$ phase under pressure. In both cases, the hexagonal phases completely decomposed into Ir and IrSb$_2$. We found that a fast cooling rate (e.g., within 1 min) was critical for obtaining the pure polycrystal. Although the present measurements are based on polycrystals, we note that both phases should be stable above $\sim$ 1200 $^{\circ}$C according to Fig. S1, which means that single crystals can be obtained for both the BHV ordering and NiAs phases.

A revised Ir-Sb phase diagram is shown in Fig. S2 that includes three main differences compared to the phase diagram in the previous report \cite{caillat1993}:\\
\textbf{a.} The vertical line representing IrSb in the intermediate temperature region has been removed where the hexagonal phase decomposes into Ir and IrSb$_2$.\\
\textbf{b.} The newly discovered $P6_3mc$ phase located at Sb = 52.94 \% (Ir$_{16}$Sb$_{18}$) is shown.\\
\textbf{c.} A solid solution region is shown in the space group of $P6_3/mmc$, namely, 50 \% $\leq$ Sb \% $<$ 52.94 \%, which is where the observed superconductivity emerges.\\

\section{Determining the structure of the parent phase of Ir$_{16}$Sb$_{18}$}

The “smoking gun” of the discovery of BHV Ir$_{16}$Sb$_{18}$ came from XRD. In contrast to previous report, the XRD patterns (Fig. S3) exhibited new forbidden peaks and spots in addition to those allowed in the NiAs structure, and these satellites could be indexed using a propagation vector of (1/3,1/3,0), indicating the formation of a new hexagonal 3$\times$3$\times$1 superlattice in the NiAs cell. Quantitatively, the diffraction peaks could be indexed to a large hexagonal lattice with $a$ = 11.889 \AA \ and $c$ = 5.525 \AA. Compared to the primary cell with $a$ = 3.97 \AA \ and $c$ = 5.53 \AA, the superstructure was tripled along the $a$-axis, while the $c$-axis remained unchanged. As per the reflection conditions of $00l$ : $l$ = 2n and $hhl$ : $l$ = 2n, we inferred three possible space groups, namely, $P6_3mc$, $P6_3/mmc$, and $P$-62$c$.

Another indication of the formation of the 3$\times$3$\times$1 superlattice can be found in the 3D electron diffraction data. A superstructure can be seen clearly from Fig. S4, for which the extinction rule is also consistent with that of the XRD data. Using the atomic-resolution images obtained through STEM observations along the $a$- and $c$-axes (Fig. 1 and Fig. S5), the only possible space configuration was determined to be $P6_3mc$. Notably, the intensity ratios of the weak/strong peaks along the $a$- [Fig. 1(e)] and $c$-axes (Fig. S5) were 2/3 and 1/2, in good agreement with the 1/3 and 1/2 vacancy occupations in each direction.

The main diffraction peaks (such as 300, 301, 302) correspond to the NiAs-primary cell with $a\sim$ 4 \AA, while the superstructure peaks (such as 100, 110, 101) correspond to a 3$\times$3 supercell with $a\sim$ 12 \AA. Considering that the BHV sample was acquired by quenching, it's reasonable to expect a more significant lattice distortion for a larger cell. This imperfection can be evidenced from the enlargement from 6-30 degrees (Fig. S3 and the inset of Fig. S6), where the superstructure peaks are systematically broadened compared with that of the main diffractions. In Fig. S6, we show the Rietveld refinement performed by using the commercial software Jade Pro 7.8.2, which could assign different shape functions for the diffraction peaks of primary cell and supercell, and the results of the refinement are summarized in Tables S1 and S2, respectively. The refined crystal structure is shown in Fig. S7, in which the surrounding six Ir and Sb atoms contract slightly towards the centre of the vacancies to relax the structure.

\section{Properties of high-pressure synthesis Ir$_{1-\delta}$Sb samples}

As the crystal structure of a material tends to be fully occupied under high-pressure synthesis conditions, we synthesized samples under different pressures (1.5–5.5 GPa) to squeeze extra Ir into the vacant sites to disrupt the BHV ordering. Here, we detail the crystal structure properties for Ir$_{1-\delta}$Sb synthesized under different pressures. As shown in Figs. S8 and S9, all the diffraction peaks of the superstructure disappeared even for the samples prepared under the lowest accessible pressure. These results suggest that the BHV ordering is suppressed when Ir fills the vacant sites. The refined parameters are summarized in Tables S1 and S2. The superconducting properties of Ir$_{1-\delta}$Sb in terms of the resistivity, magnetic susceptibility, and specific heat are shown in Fig. S10. Note that the normalized specific-heat jump value $\Delta C$/$\gamma T_c$ = 1.94 for the sample synthesized under 1.5 GPa (Fig. S11) is larger than the Bardeen–Cooper–Schrieffer (BCS) expected value of 1.43, indicates the possible underlying strong coupling mechanisms such as strong electron-phonon coupling, charge fluctuation, etc.

\section{Properties of Ir$_{1-x}$Rh$_x$Sb}

A series of Ir$_{1-x}$Rh$_x$Sb samples were synthesized via arc melting at ambient pressure. From the XRD patterns of all Ir$_{1-x}$Rh$_x$Sb samples [Fig. S12(a)], we observed a large solid solution of Rh up to 60 \%. As the Rh content further increased, a first-order phase transition from the NiAs- to MnP-type was observed. The Rietveld refinement of Rh = 3\% ($P6_3mc$) and 50\% ($P6_3/mmc$) are shown in Fig. S12(b) and (c). An SEM image for the Rh-doped sample is also shown in Fig. S13. The composite image together with the continuous shift of the XRD peak [Fig. 2(c)] suggests the homogeneity of the Rh substitution. Finally, magnetic data are shown for the Ir$_{1-x}$Rh$_x$Sb samples to confirm the superconducting transition (Fig. S14). The superconducting volume fractions of Ir$_{1-x}$Rh$_x$Sb are summarized here together with the evolution of the degree of disorder (Fig. S15). Bulk superconductivity was achieved when the BHV ordering was completely suppressed.

\section{results of the calculation}

Here, we detail the band structure and Fermi surface topology of Ir$_{1-\delta}$Sb (Figs. S16 and S17). The series of vacancy calculations (Figs. S18–S20) were based on a framework of a 3$\times$3$\times$1 supercell according to the Fermi surface nesting vector of (1/3, 1/3, 0). Otherwise, an infinite number of possible configurations would exist. Thus, we propose the synergistic effect of the Fermi surface nesting and vacancy formation energy as a possible origin for the experimentally observed BHV ordering structure. Fig. S21 illustrates the full range of unfolded bands for various vacancy arrangements. Compared to other configurations, the BHV ordering can mostly smear the bands and transfer the electron spectral weight near the Fermi surface.

~\\
\setcounter{figure}{0}
\makeatletter 
\renewcommand{\thefigure}{S\@arabic\c@figure}
\makeatother

\pagebreak

\begin{table*}[h]
\renewcommand{\thetable}{S\arabic{table}}
\caption{\label{tab:table1}
Refinement results of Ir$_{16}$Sb$_{18}$ and Ir$_{1-\delta}$Sb}
\hrule width \hsize \kern 1mm \hrule width \hsize \kern 1mm
\begin{tabular}{cccc}
 & Ir$_8$Sb$_9$ (Ir$_{16}$Sb$_{18}$) & Ir$_{1-\delta}$Sb (1.5 GPa) & Ir$_{1-\delta}$Sb (5.5 GPa)\\
 \hline
molecular weight & 2633.576 & 298.59964 & 313.977\\
formula (EPMA) & Ir$_8$Sb$_9$ & Ir$_{0.92}$Sb & Ir$_{0.965}$Sb\\
lattice & hexagonal & hexagonal & hexagonal\\
space group & $P6_3mc$ & $P6_3/mmc$ & $P6_3/mmc$\\
$a$ (\AA) & 11.92723(4) & 3.99544(4) & 4.00775(2)\\
$c$ (\AA) & 5.53375(3) & 5.51413(6) & 5.51784(6)\\
V (\AA$^3$) & 678.529350 & 76.231922 & 76.753993\\
$Z$ & 2 & 2 & 2\\
$T$ (K) & 298(2) & 298(2) & 298(2)\\
radiation type & Cu & Cu & Cu\\
R$_p$ (\%) & - & 9.34 & 7.7\\
R$_{wp}$ (\%) & 5.84 & 12.6 & 10.3\\
R$_{exp}$ (\%) & 3.60 & 7.37 & 7.28\\
$\chi^2$ & 2.62 & 2.93 & 2.0\\
\end{tabular}
\hrule width \hsize \kern 1mm \hrule width \hsize \kern 1mm
\end{table*}

\begin{table*}[h]
\renewcommand{\thetable}{S\arabic{table}}
\caption{\label{tab:table2}
Atomic locations of Ir$_{16}$Sb$_{18}$ and Ir$_{1-\delta}$Sb}
\hrule width \hsize \kern 1mm \hrule width \hsize \kern 1mm
\begin{tabular}{cccccc}
 & atom & x & y & z & occ.\\
 \hline
& Ir$_1$ & -0.01091(19) & 0.33022(25) & 0.02123(114) & 0.991\\
& Ir$_2$ & 0.0 & 0.0 & 0.02055(296) & 0.980\\
Ir$_8$Sb$_9$ & Ir$_3$ & 0.33333 & 0.66667 & 0.02663(311) & 0.96\\
(Ir$_{16}$Sb$_{18}$) & Sb$_1$ & 0.11199(63) & 0.22394(126) & 0.24466(105) & 0.996\\
& Sb$_2$ & 0.11478(70) & 0.55739(35) & 0.25548(109) & 0.971\\
& Sb$_3$ & 0.45274(64) & 0.22637(32) & 0.26876(104) & 1.0\\
\hline
Ir$_{1-\delta}$Sb  & Ir & 0 & 0 & 0 & 0.966\\
(5.5 GPa) & Sb & 0.3333 & 0.6667 & 0.25 & 1.0\\
\end{tabular} 
\hrule width \hsize \kern 1mm \hrule width \hsize \kern 1mm
\end{table*}

\begin{table*}[h]
\renewcommand{\thetable}{S\arabic{table}}
\caption{\label{tab:table2}
Atomic locations of Ir$_{0.97}$Rh$_{0.03}$Sb and Ir$_{0.5}$Rh$_{0.5}$Sb}
\hrule width \hsize \kern 1mm \hrule width \hsize \kern 1mm
\begin{tabular}{cccccc}
 & atom & x & y & z & occ.\\
 \hline
& Ir(Rh)$_1$ & 0.0 & 0.08333(12) & 0.14503(301) & -\\
& Ir(Rh)$_2$ & -0.00590(331) & 0.3319(14) & 0.0949(58) & -\\
Ir$_{0.864}$Rh$_{0.028}$Sb (EPMA) & Ir(Rh)$_3$ & 0.33333 & 0.66667 & 0.1273(19) & -\\
($a$=11.9286(1)\AA, $c$=5.5338(2)\AA) & Sb$_1$ & 0.1094(67) & 0.2187(60) & 0.3339(205) & -\\
 & Sb$_2$ & 0.1170(45) & 0.5585(34) & 0.3427(41) & -\\
& Sb$_3$ & 0.4529(11) & 0.2265(7) & 0.3380(21) & -\\
\hline
 & Ir & 0 & 0 & 0 & 0.44\\
Ir$_{0.5}$Rh$_{0.5}$Sb & Rh & 0 & 0 & 0 & 0.45\\
($a$=3.9653(1)\AA, $c$=5.5615(2)\AA) & Sb & 0.3333 & 0.6667 & 0.25 & 0.94\\
\end{tabular} 
\hrule width \hsize \kern 1mm \hrule width \hsize \kern 1mm
\end{table*}

\begin{figure*}[t]%
\centering
\includegraphics[width=12cm]{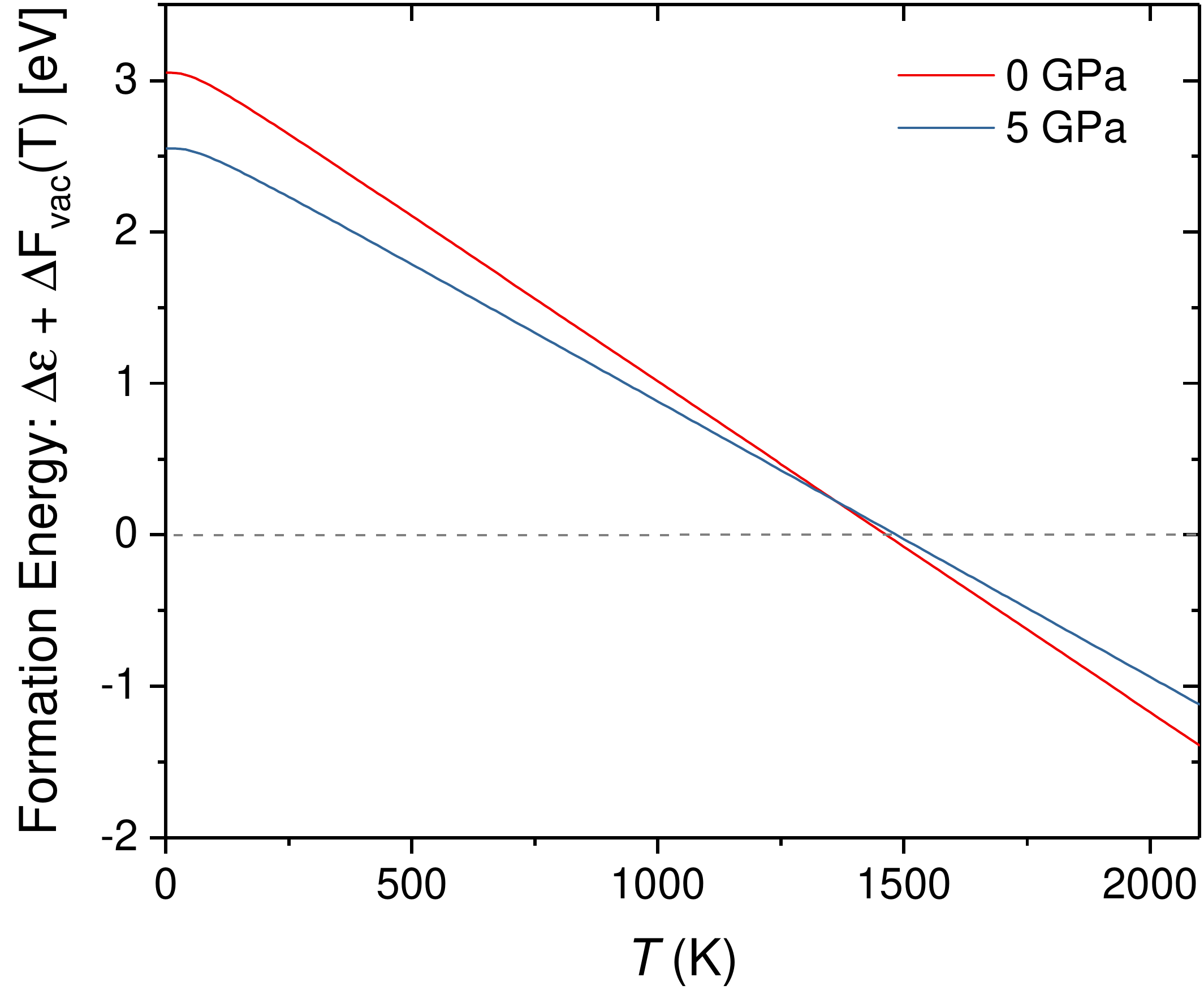}
\caption{\label{figs1} Formation energy of the vacancies contained in Ir$_{16}$Sb$_{18}$ at ambient pressure and in fully occupied IrSb at 5 GPa.}
\end{figure*}

\begin{figure*}[t]%
\centering
\includegraphics[width=16cm]{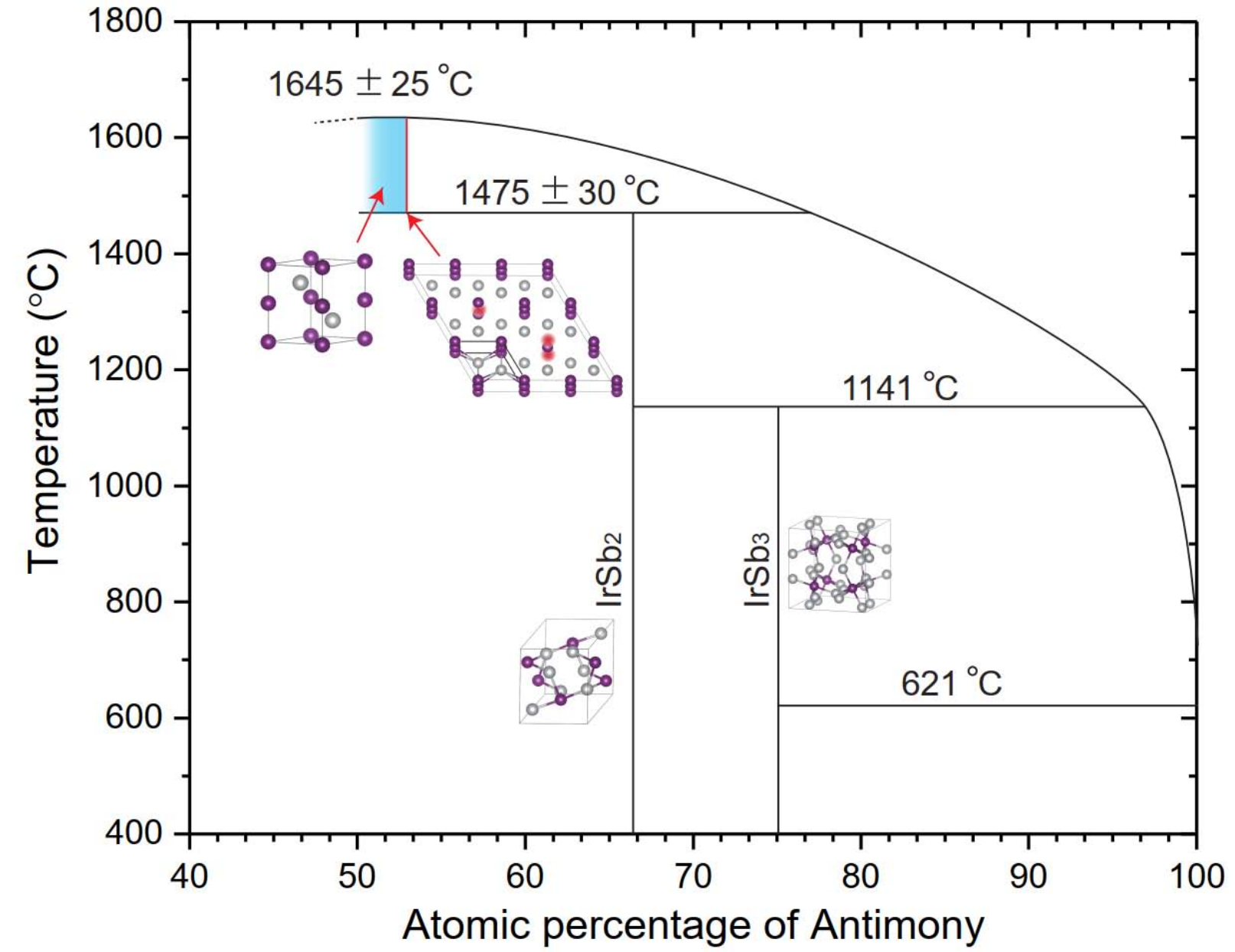}
\caption{\label{figs2} Revised phase diagram of Ir-Sb binary compounds. The temperatures in the figure were adopted without modification from Ref. \cite{caillat1993}. The corresponding crystal structures are superimposed on the figure. The purple and grey circles represent the Ir and Sb atoms, respectively. The red vertical line indicates the newly discovered $P6_3mc$ (Ir$_{16}$Sb$_{18}$), while the blue region represents a solid solution of Ir$_{1-\delta}$Sb in the space group of $P6_3/mmc$.}
\end{figure*}

\begin{figure*}[t]%
\centering
\includegraphics[width=14cm]{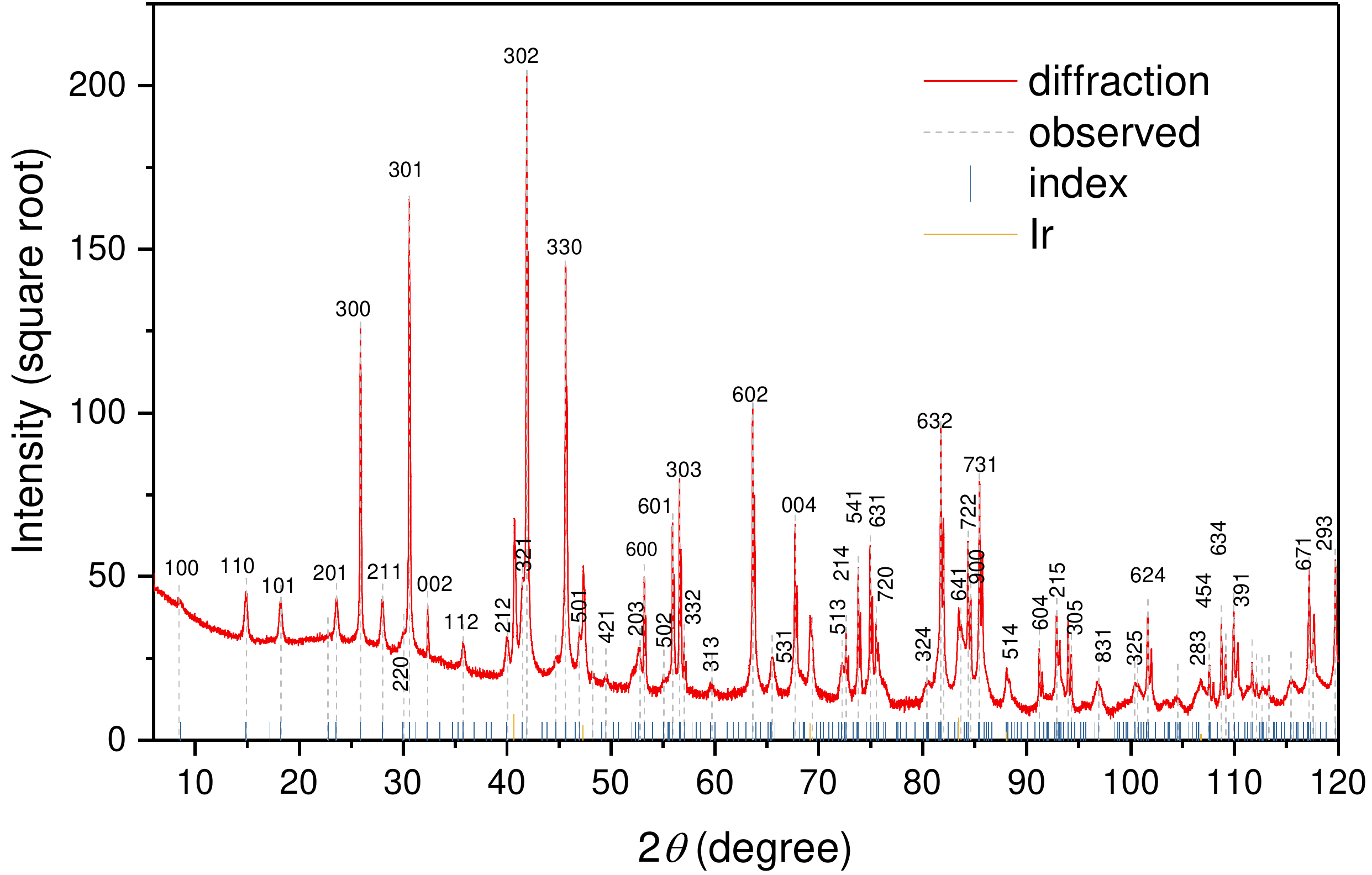}
\caption{\label{figs3} XRD pattern of the BHV-ordering phase. The intensity is depicted using the square root scale to show the superstructure diffraction peaks more clearly. The Laue indices are labelled.}
\end{figure*}

\begin{figure*}[t]%
\centering
\includegraphics[width=12cm]{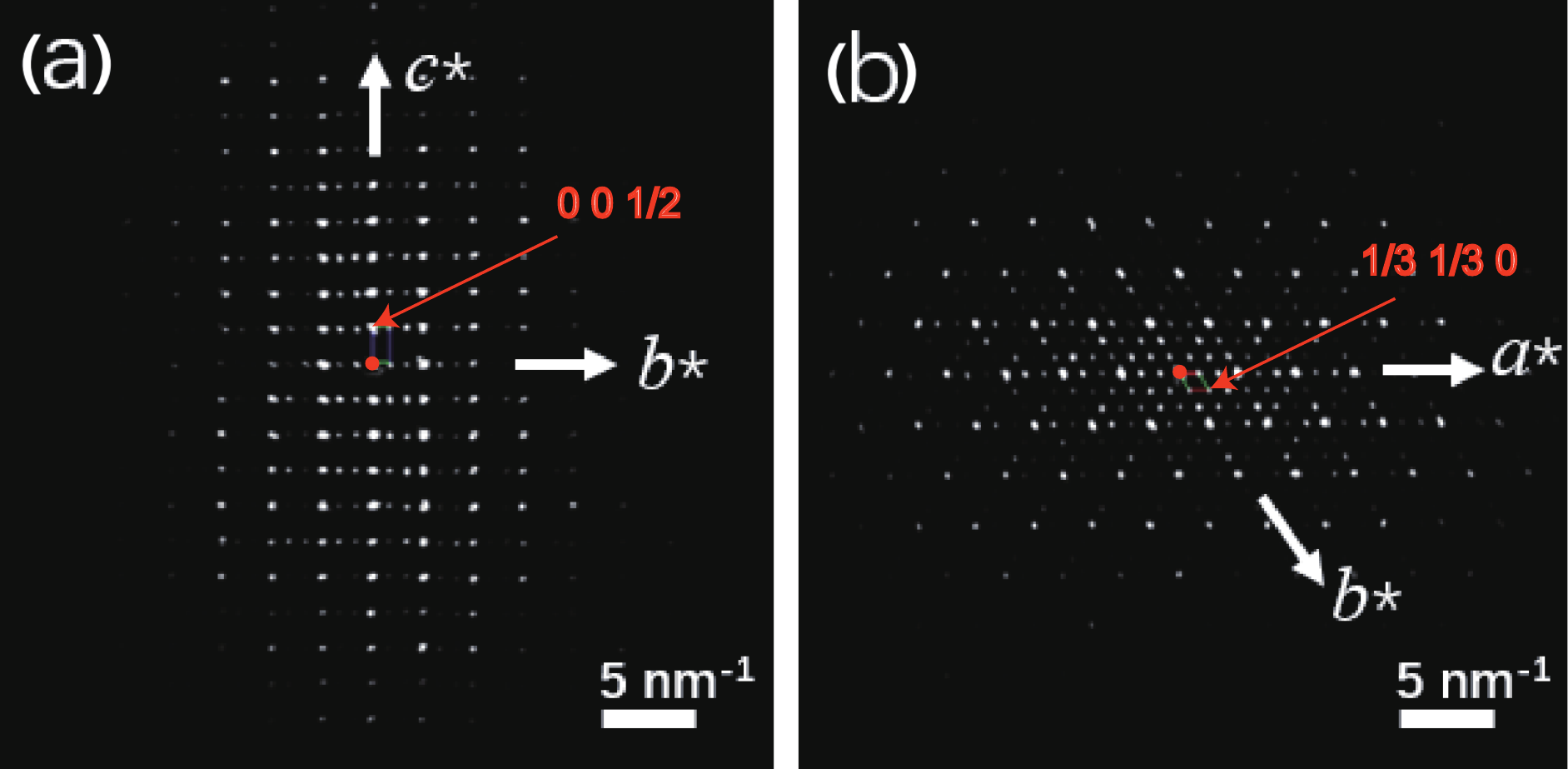}
\caption{\label{figs4} Electron diffraction data of Ir$_{16}$Sb$_{18}$. (a) and (b) show the projections of the 3D dataset along the $a$- and $c$-axes, respectively.}
\end{figure*}

\begin{figure*}[t]%
\centering
\includegraphics[width=16cm]{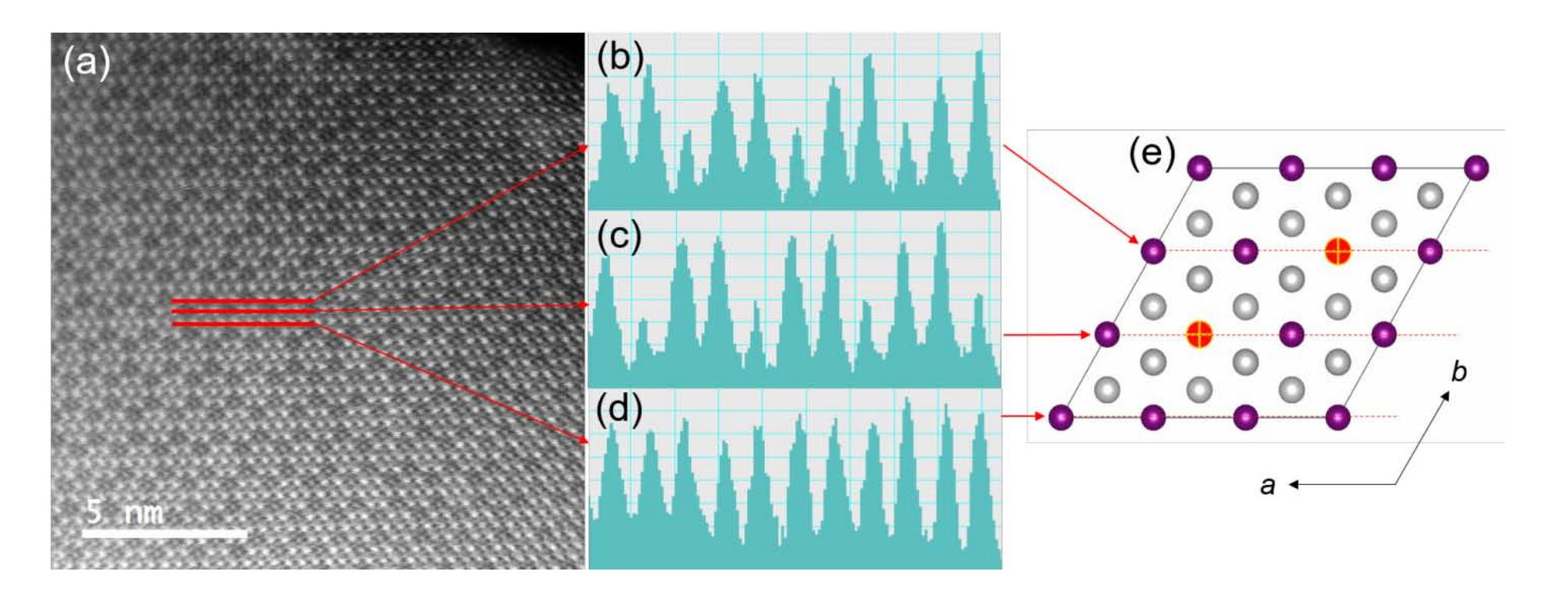}
\caption{\label{figs5} (a). HAADF micrograph of Ir$_{16}$Sb$_{18}$ along the $c$-axis and its intensity line profiles (b)-(d) of the regions indicated by arrows. The ratio of the weak/strong peaks shown in (b) and (c) is 1/2 after subtracting the background due to the half occupation of iridium atoms in the projection along the $c$-axis. (e). Projection of the BHV superstructure along the $c$-axis. The red, purple, and grey circles represent vacancies, iridium, and antimony, respectively.}
\end{figure*}

\begin{figure*}[t]%
\centering
\includegraphics[width=12cm]{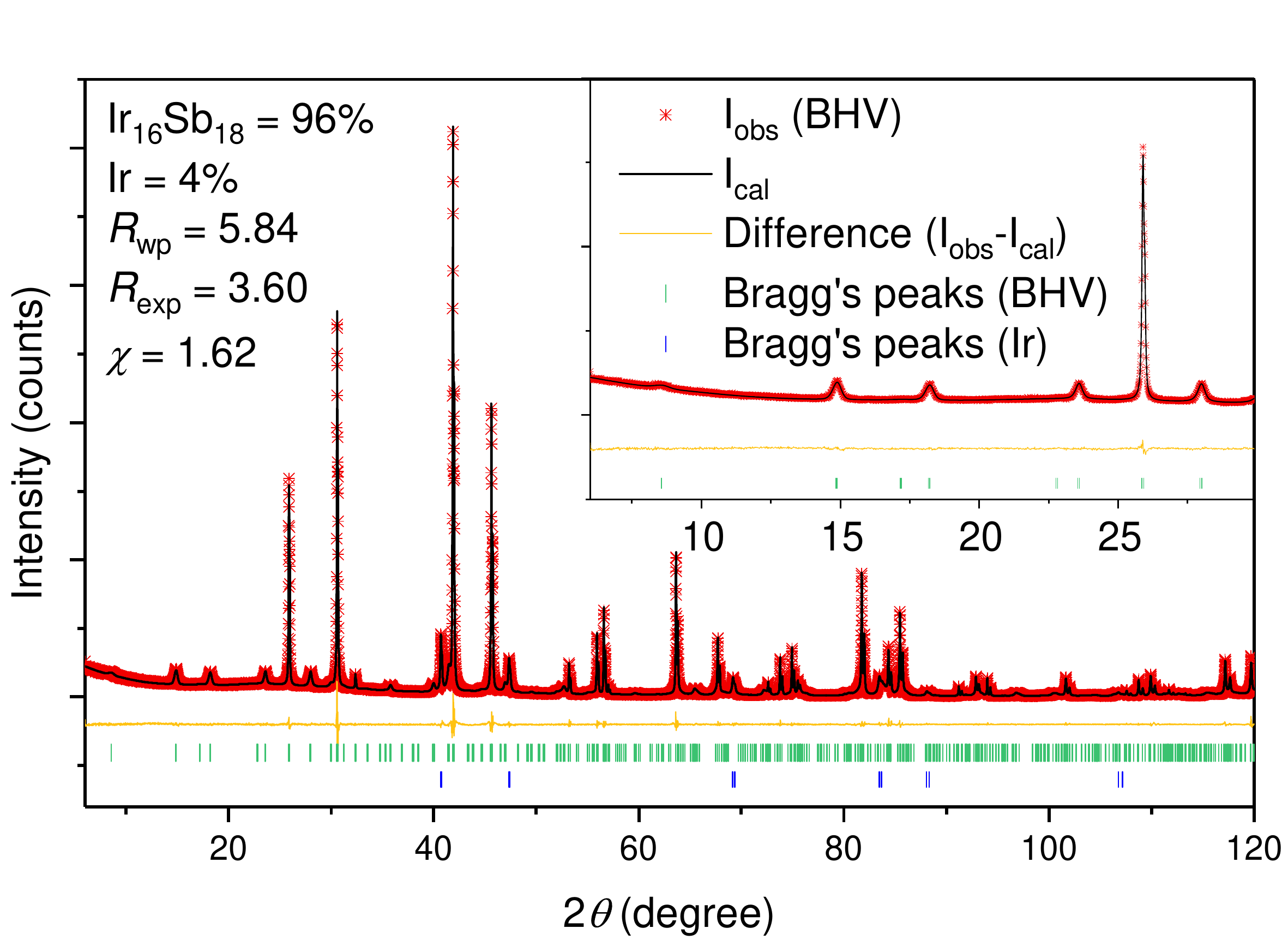}
\caption{\label{figs6} Rietveld refinement of Ir$_{16}$Sb$_{18}$. Inset is the enlargement from 6$^\circ$-30$^\circ$.}
\end{figure*}

\begin{figure*}[t]%
\centering
\includegraphics[width=10cm]{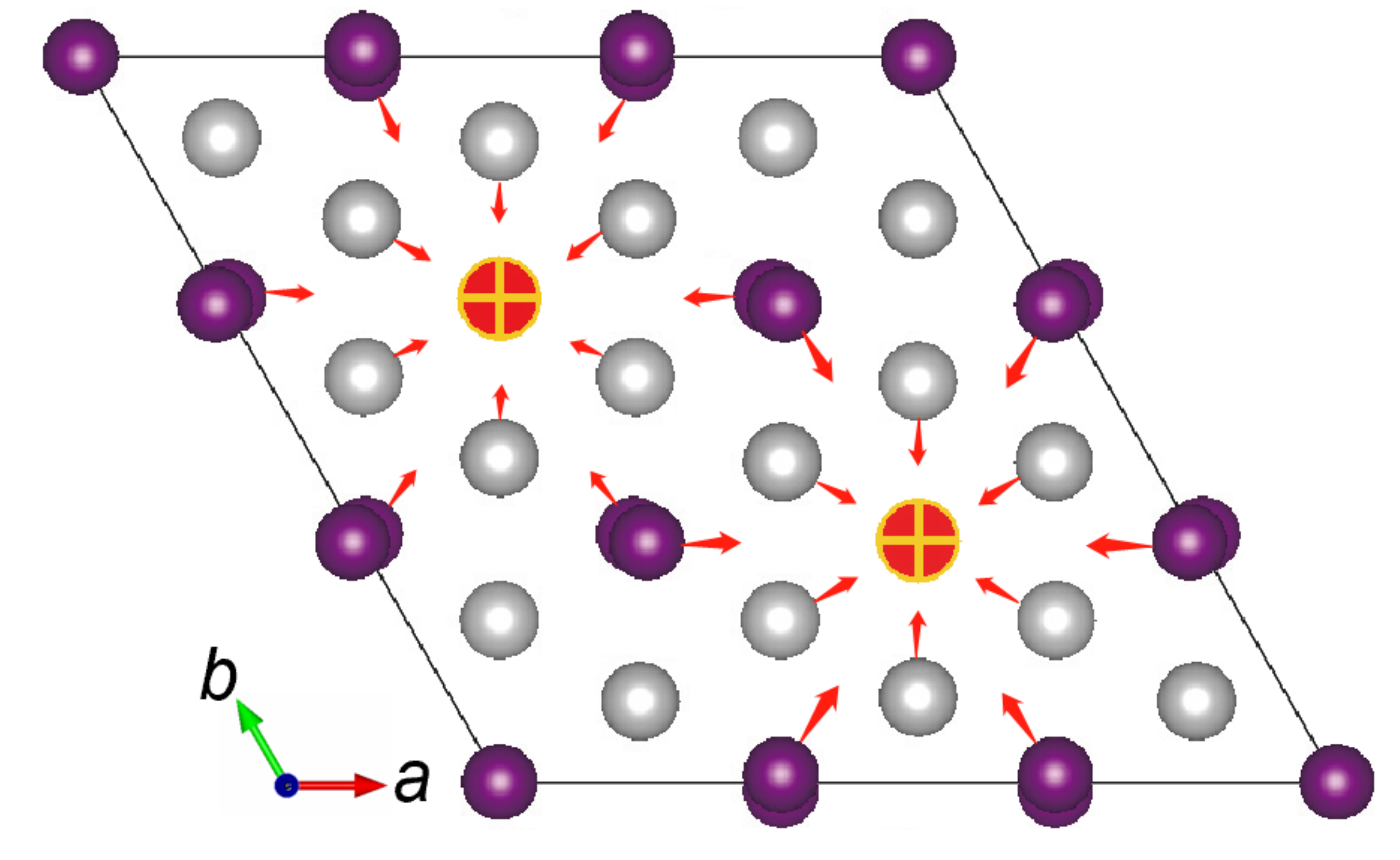}
\caption{\label{figs7} Crystal structure of Ir$_{16}$Sb$_{18}$ after refinement. The presence of iridium vacancies causes the surrounding six iridium/antimony atoms to contract slightly towards the vacancy centre to relax the structure. The red, purple, and grey circles represent vacancies, iridium, and antimony, respectively.}
\end{figure*}

\begin{figure*}[t]%
\centering
\includegraphics[width=12cm]{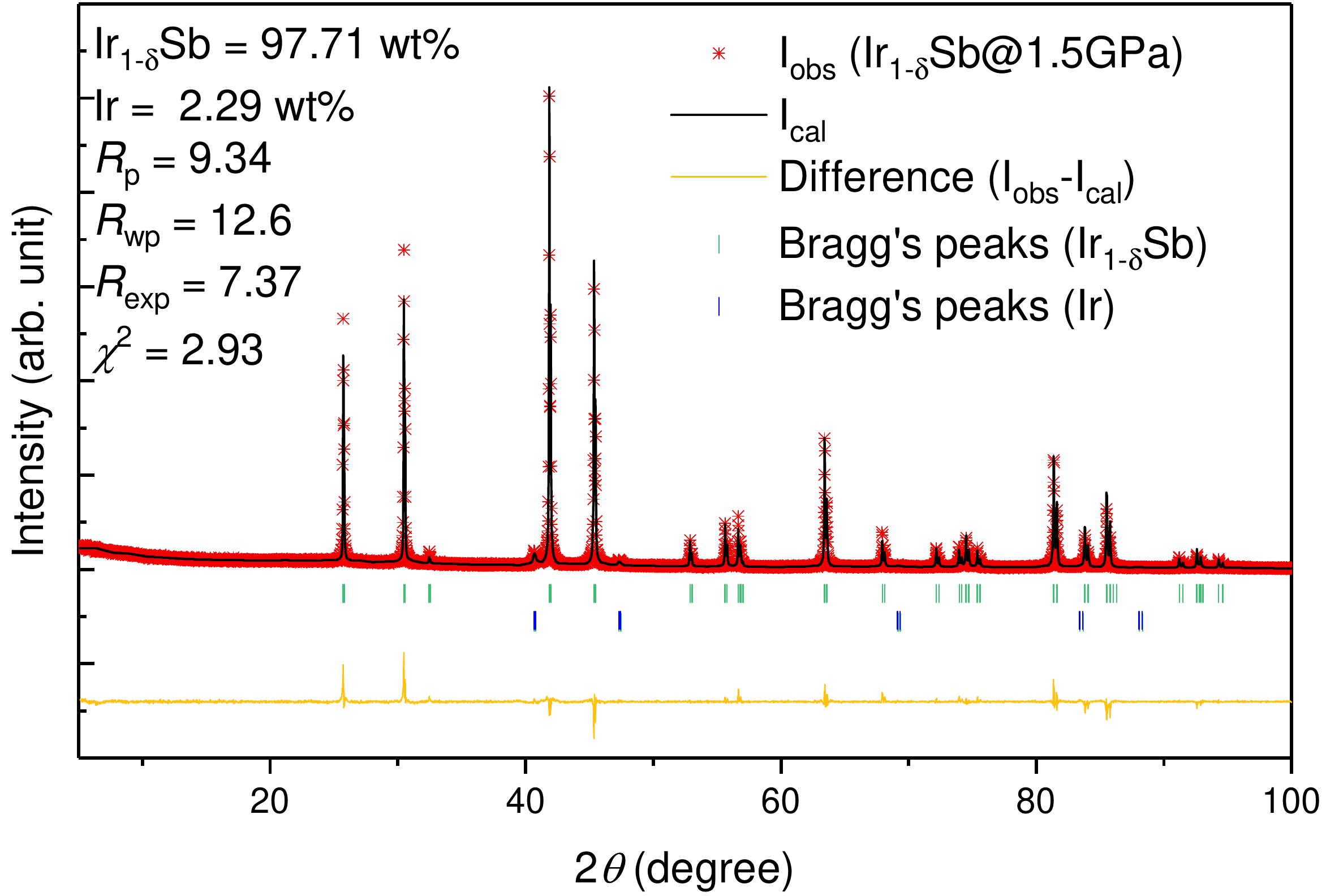}
\caption{\label{figs8} Rietveld refinement of Ir$_{1-\delta}$Sb synthesized at 1.5 GPa. Compared to Ir$_{16}$Sb$_{18}$, all of the diffraction peaks of the superstructure disappeared, which indicates successful filling of Ir into the vacant sites and transformation of the superstructure into a simple NiAs primary cell. The refined parameters are listed in Tables S1 and S2.}
\end{figure*}

\begin{figure*}[t]%
\centering
\includegraphics[width=12cm]{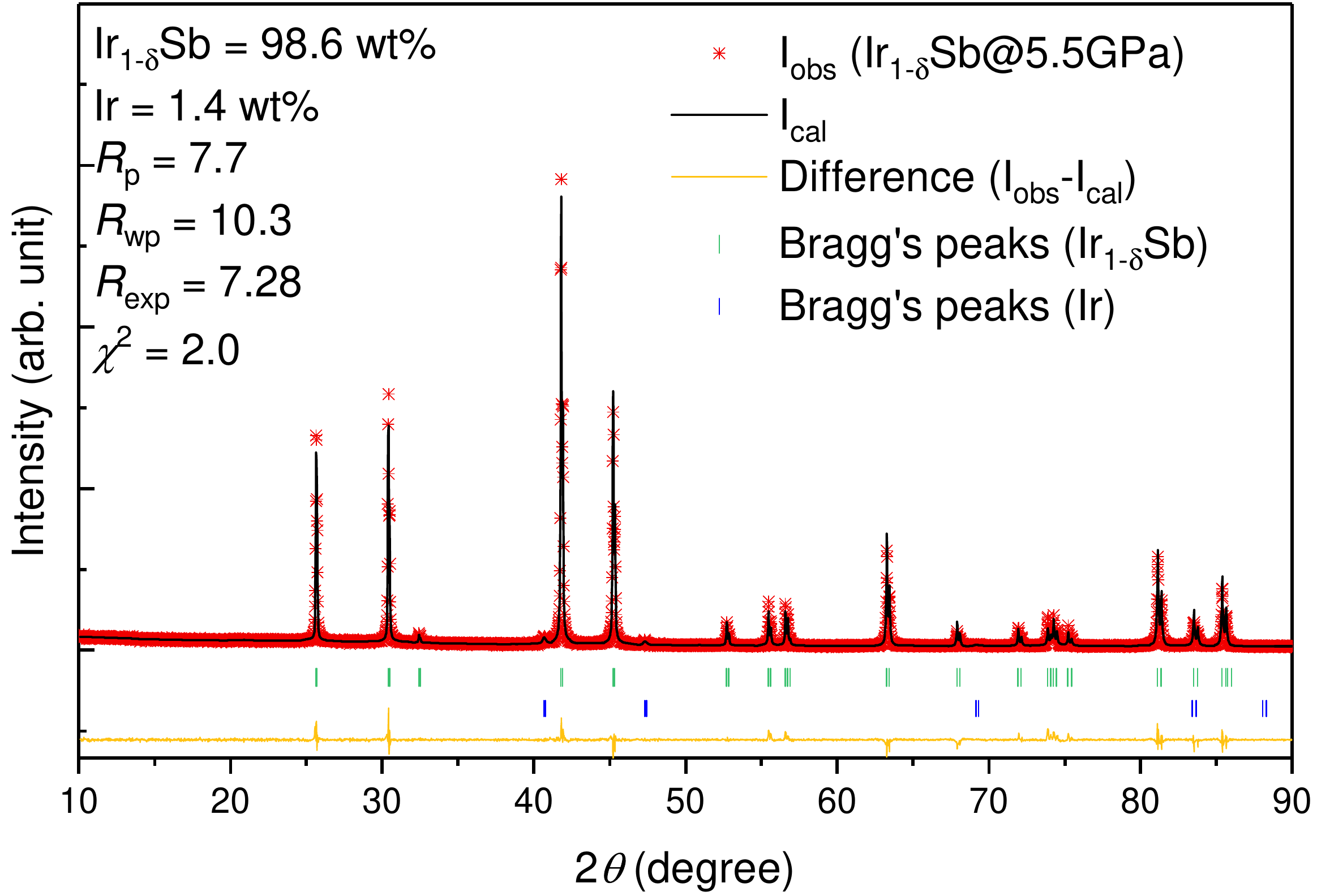}
\caption{\label{figs9} Rietveld refinement of Ir$_{1-\delta}$Sb synthesized at 5.5 GPa. The refined parameters are listed in Tables S1 and S2.}
\end{figure*}

\begin{figure*}[t]%
\centering
\includegraphics[width=\textwidth]{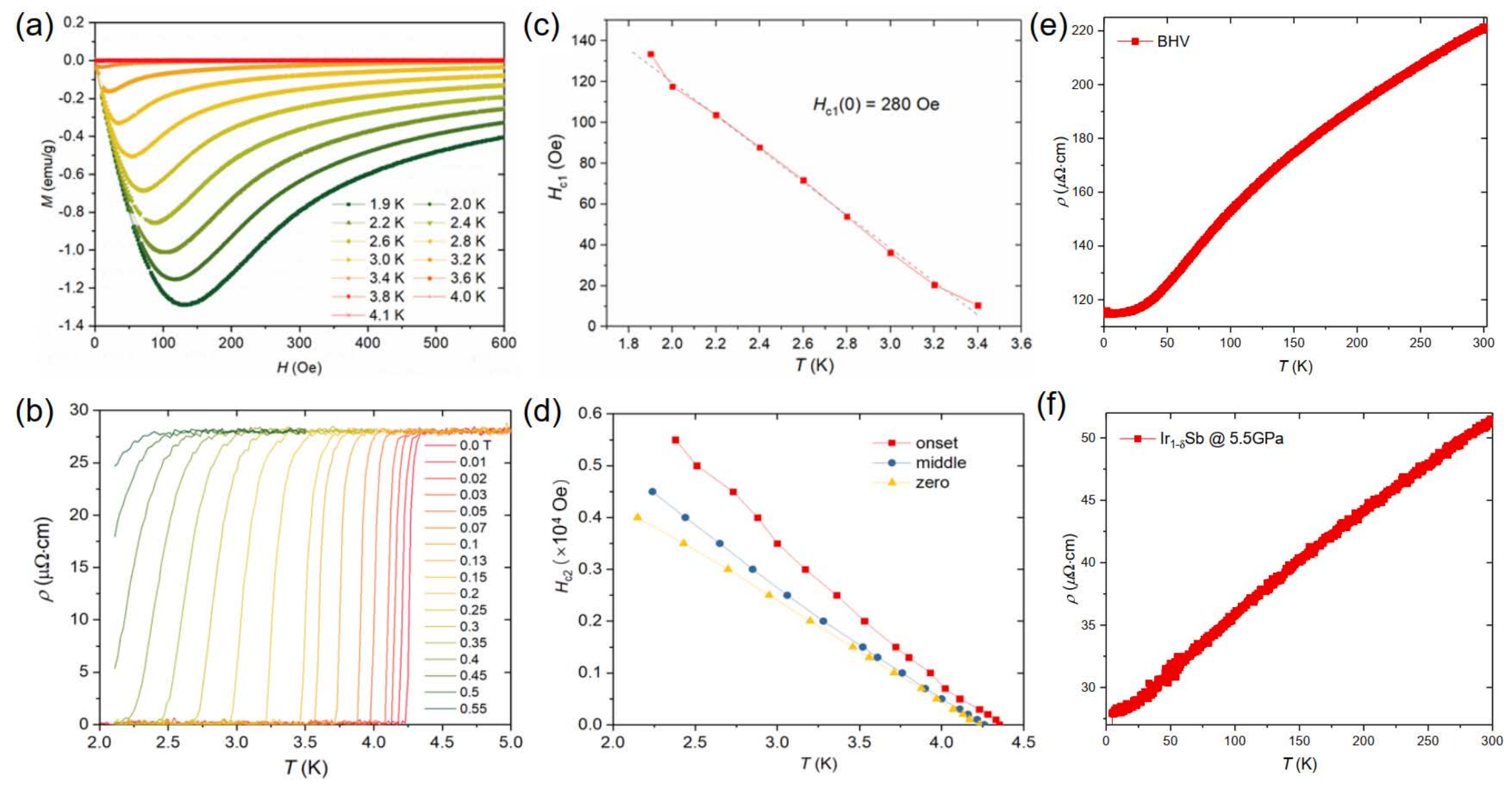}
\caption{\label{figs10} (a) Isothermal magnetization of Ir$_{1-\delta}$Sb synthesized at 5.5 GPa. (b) Lower critical field $H_{c1}$ as a function of temperature. The dotted line is a linear fit for the relation of $\mu_0H_{c1}$(T) = $\mu_0H_{c1}(0)[1 - (T/T_c)^2$]. The value of $\mu_0H_{c1}$(0) was fitted to 280 Oe. (c) Temperature dependence of the resistivity for Ir$_{1-\delta}$Sb under different magnetic fields. (d) Upper critical field $\mu_0H_{c2}$(0) as a function of temperature. By using the Werthamer-Helfand-Hohenberg formula, we obtain $\mu_0H_{c2}$(0) values of 1.19 T, 0.93 T, and 0.82 T for $T_c^{onset}$, $T_c^{middle}$, and $T_c^{zero}$, respectively. (e) and (f) The normal state resistivity of BHV ordering and 5.5 GPa samples from 2 to 300 K.}
\end{figure*}

\begin{figure*}[t]%
\centering
\includegraphics[width=10cm]{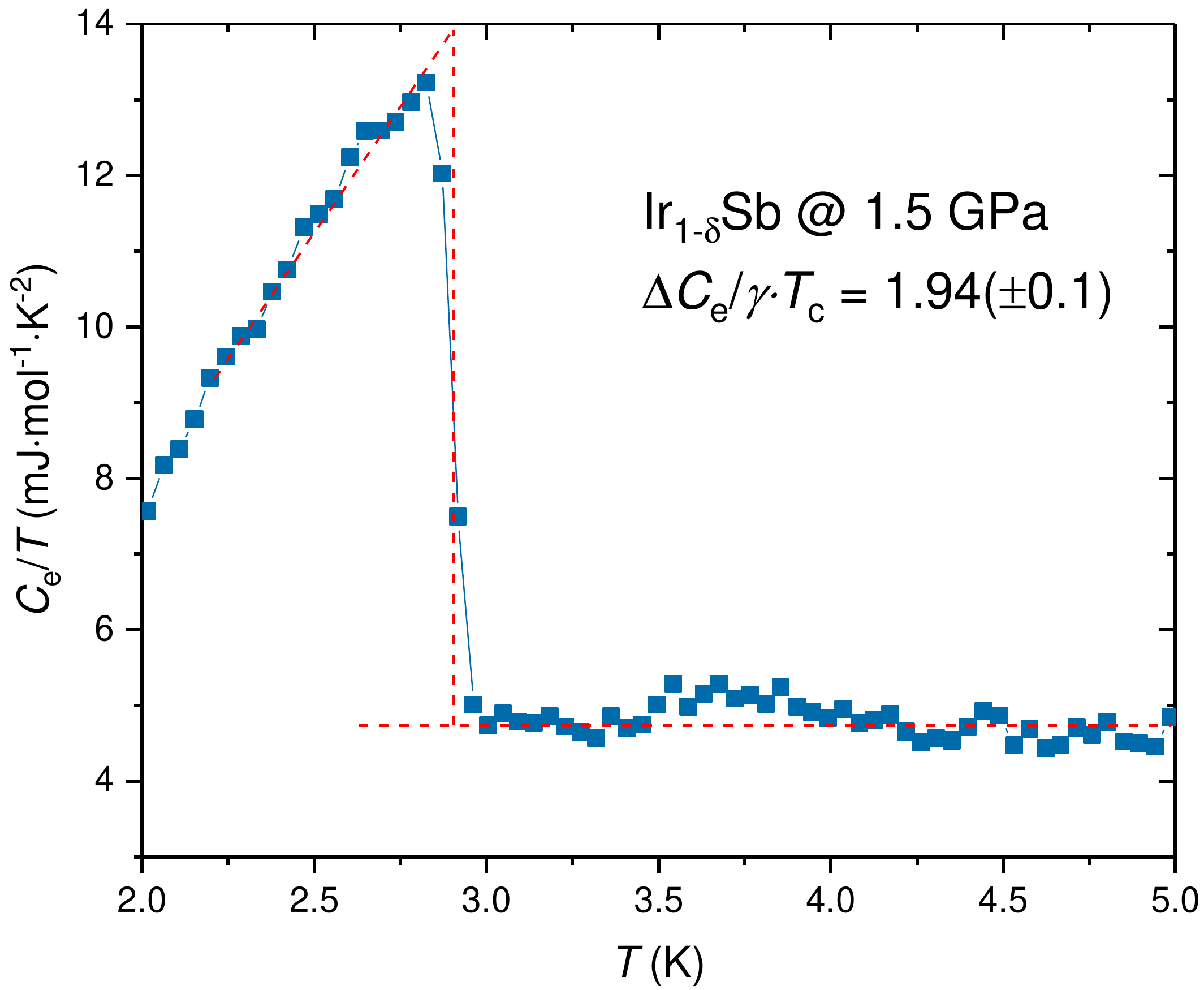}
\caption{\label{figs11} Electronic specific heat ($C_e/T$) of Ir$_{1-\delta}$Sb synthesized under 1.5 GPa. We fitted the normal-state heat capacity data under 1 T to suppress the superconductivity using the equation $C_p/T$ = $\gamma$ + $\beta T^2$, where $\gamma$ is the Sommerfeld coefficient and $\beta$ is the phonon contribution. The phonon part was subtracted from the zero field data. The normalized specific-heat jump value $\Delta C/\gamma T_c$ = 1.94 is larger than the Bardeen–Cooper–Schrieffer (BCS) weak coupling value of 1.43.}
\end{figure*}

\begin{figure*}[t]%
\centering
\includegraphics[width=10cm]{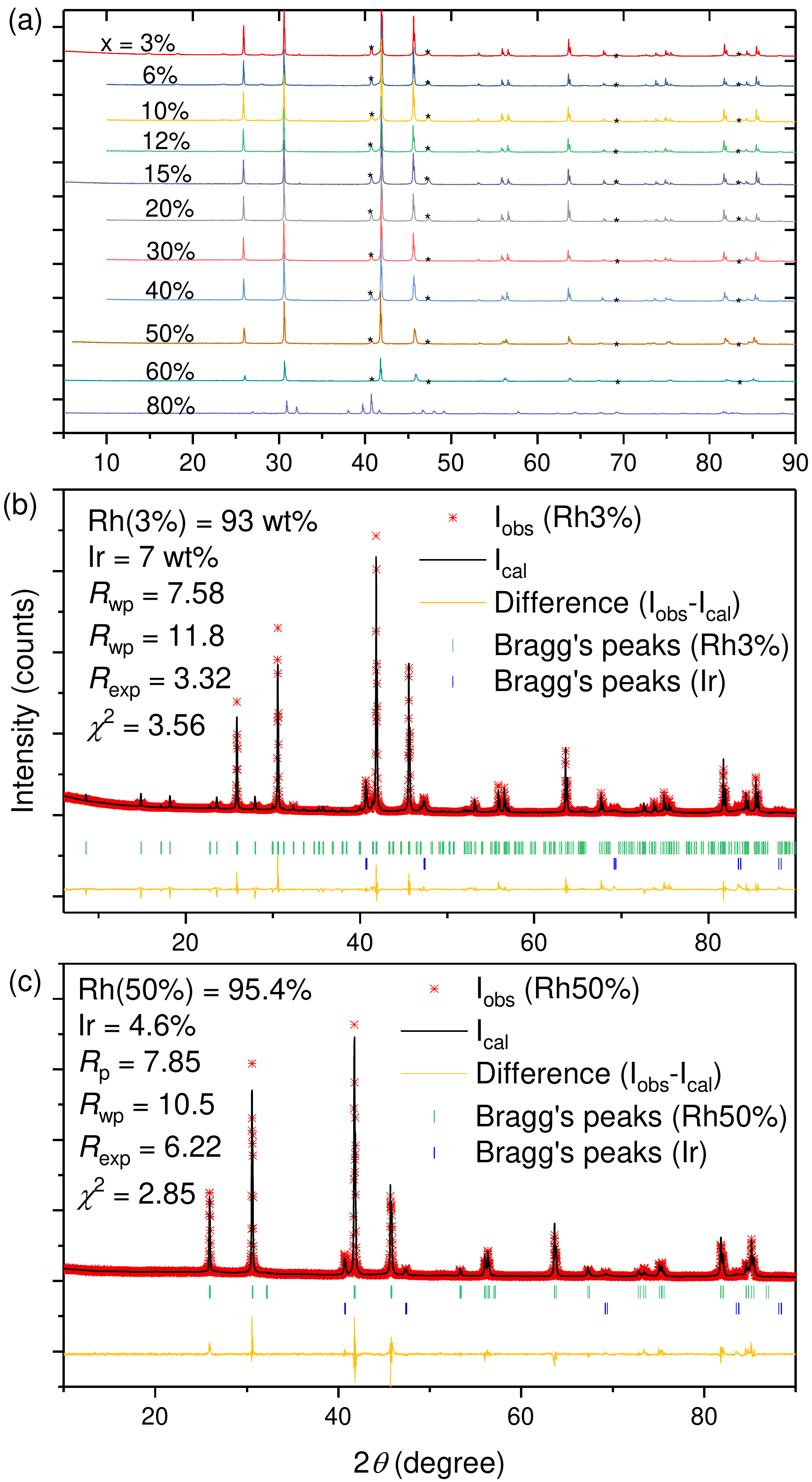}
\caption{\label{figs12} (a) XRD patterns for the Ir$_{1-x}$Rh$_x$Sb samples. The NiAs-type structure was preserved up to $x$ = 60 \% and was completely transformed into the MnP structure by $x$ = 80 \%. The low vapour pressure of Sb compared to that of Ir during the arc melting process made the loss of Sb unavoidable. The stars indicate small amounts of residual Ir. (b) and (c) Rietveld refinement of Rh = 3\% ($P6_3mc$) and 50\% ($P6_3/mmc$) samples.}
\end{figure*}

\begin{figure*}[t]%
\centering
\includegraphics[width=12cm]{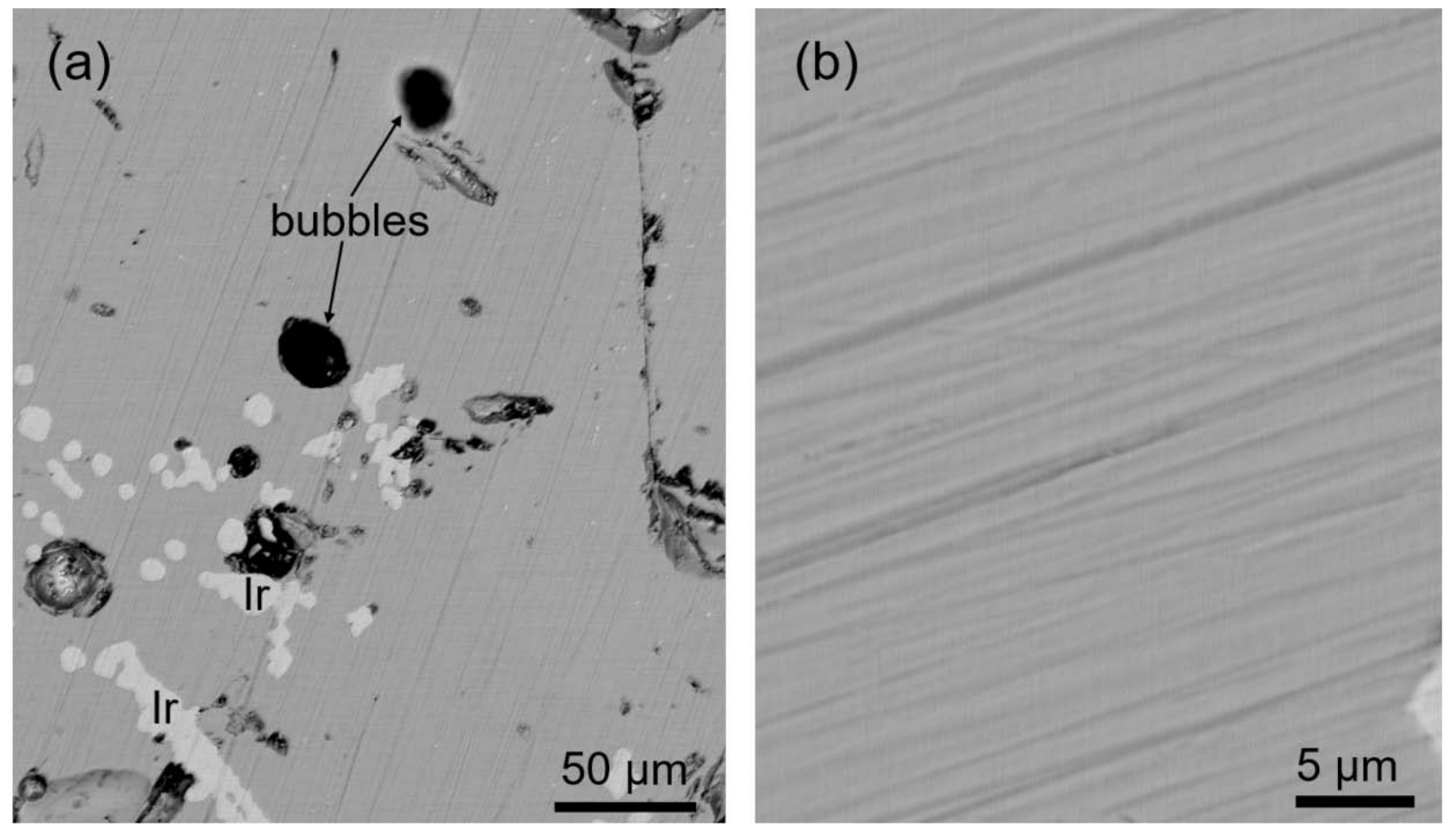}
\caption{\label{figs13} (a),(b) Low and high magnification SEM images of Ir$_{0.97}$Rh$_{0.03}$Sb. The surface of the sample was polished before the measurements. The bubbles and residual Ir formed during the arc melting due to the low vapour pressure of Sb.}
\end{figure*}

\begin{figure*}[t]%
\centering
\includegraphics[width=12cm]{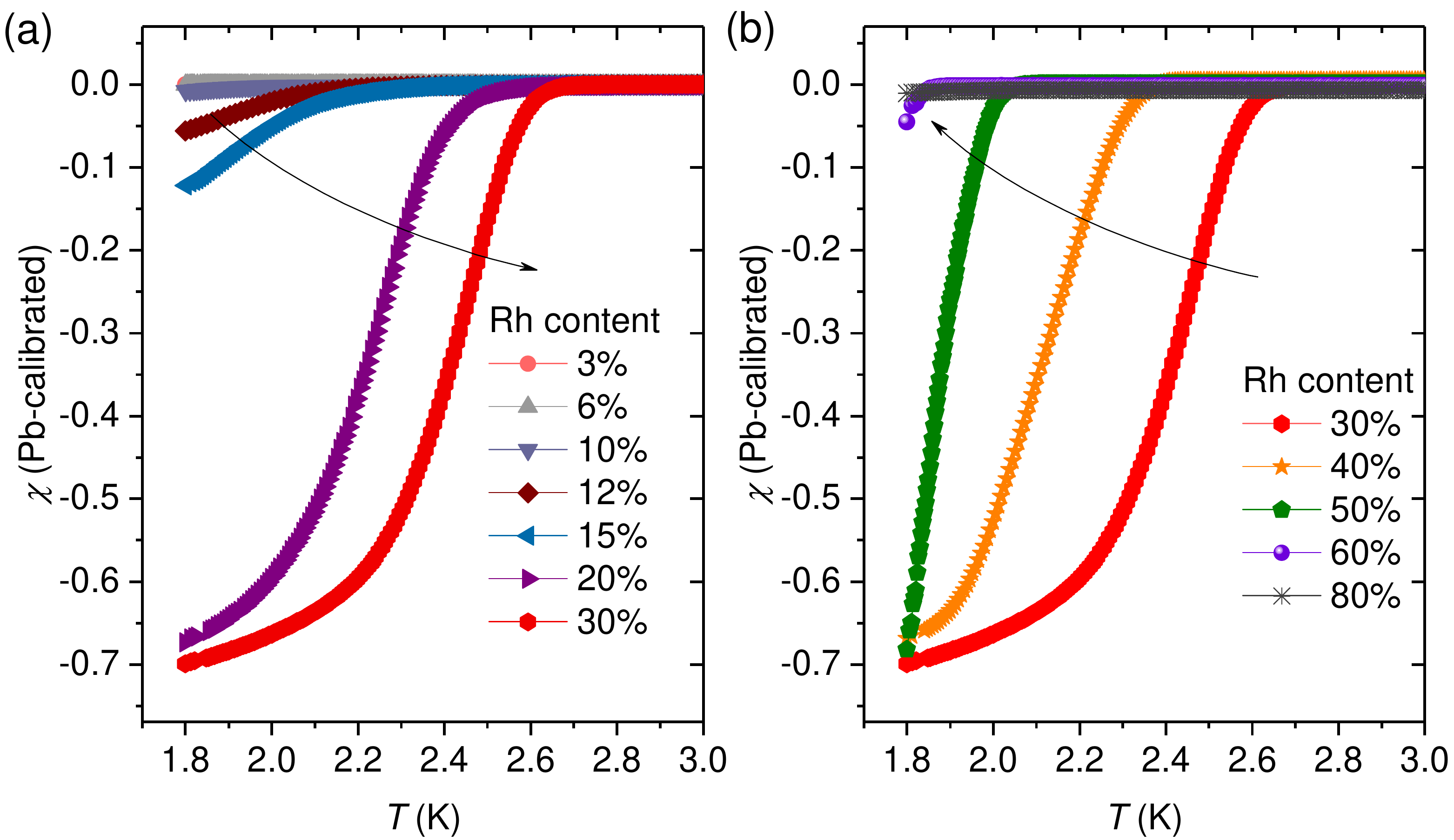}
\caption{\label{figs14} Diamagnetization of Ir$_{1-x}$Rh$_x$Sb. All samples were calibrated with reference to Pb as a standard.}
\end{figure*}

\begin{figure*}[t]%
\centering
\includegraphics[width=12cm]{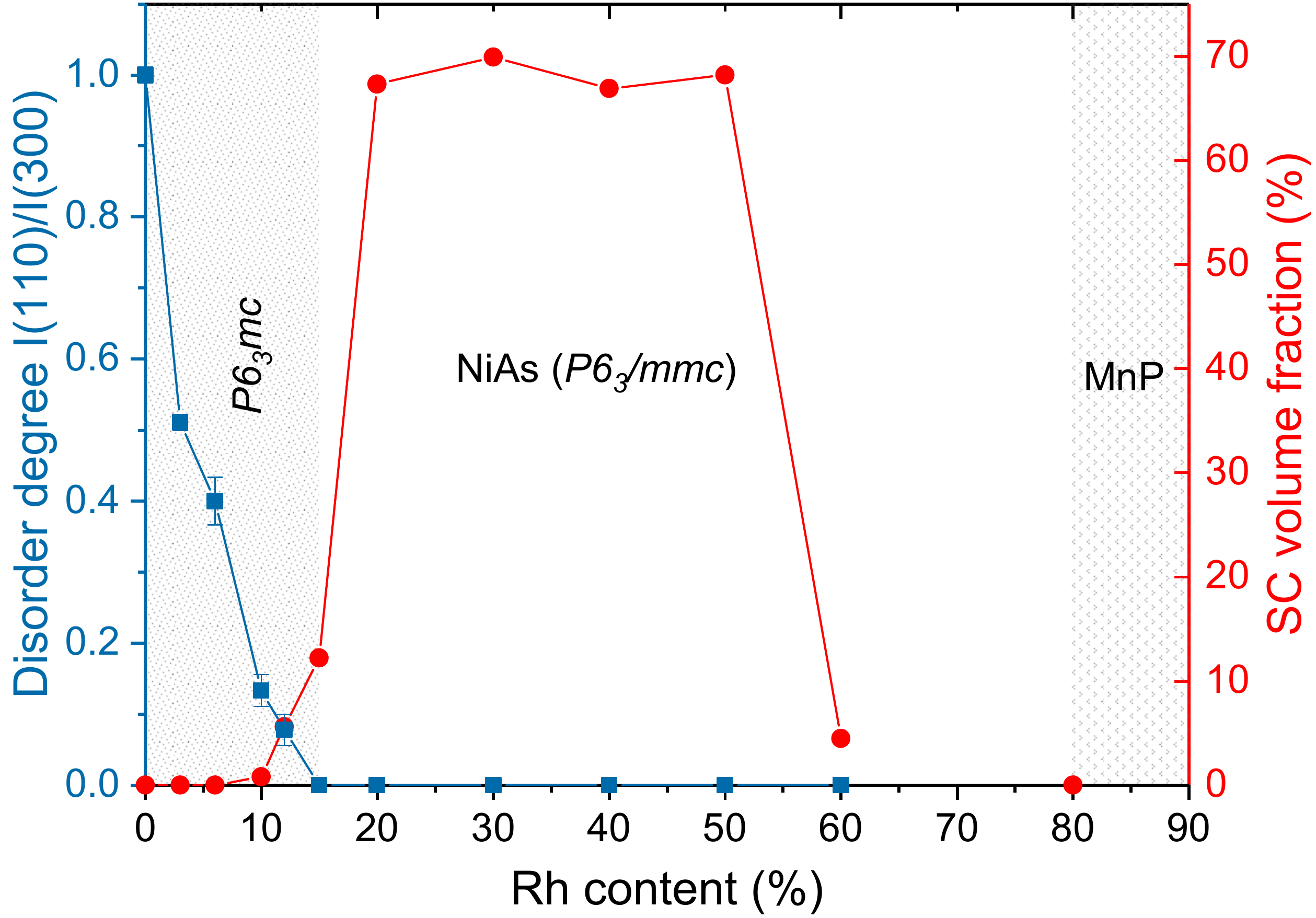}
\caption{\label{figs15} Superconducting volume fraction of Ir$_{1-x}$Rh$_x$Sb from Fig. S14. The evolution of the degree of disorder is also shown. The order degree is defined as the normalized peak ration of (110/300) with the reference to that of the Ir$_{16}$Sb$_{18}$ ($x$ = 0 \%)}
\end{figure*}

\begin{figure*}[t]%
\centering
\includegraphics[width=15cm]{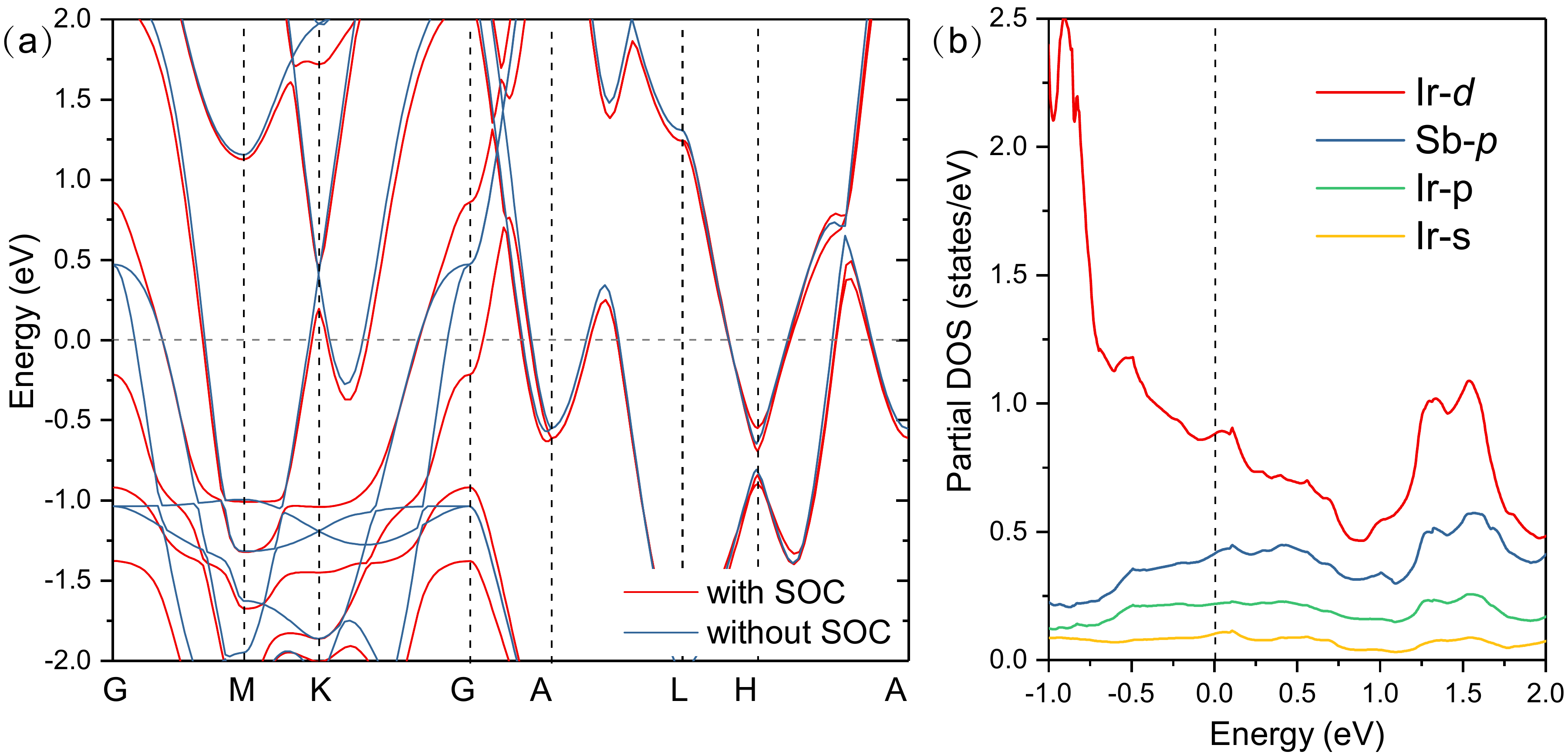}
\caption{\label{figs16} (a) Band dispersion of pristine IrSb with and without SOC. (b) Partial DOS of IrSb. The Fermi level is dominated by Ir-$d$ and Sb-$p$ electrons.}
\end{figure*}

\begin{figure*}[t]%
\centering
\includegraphics[width=12cm]{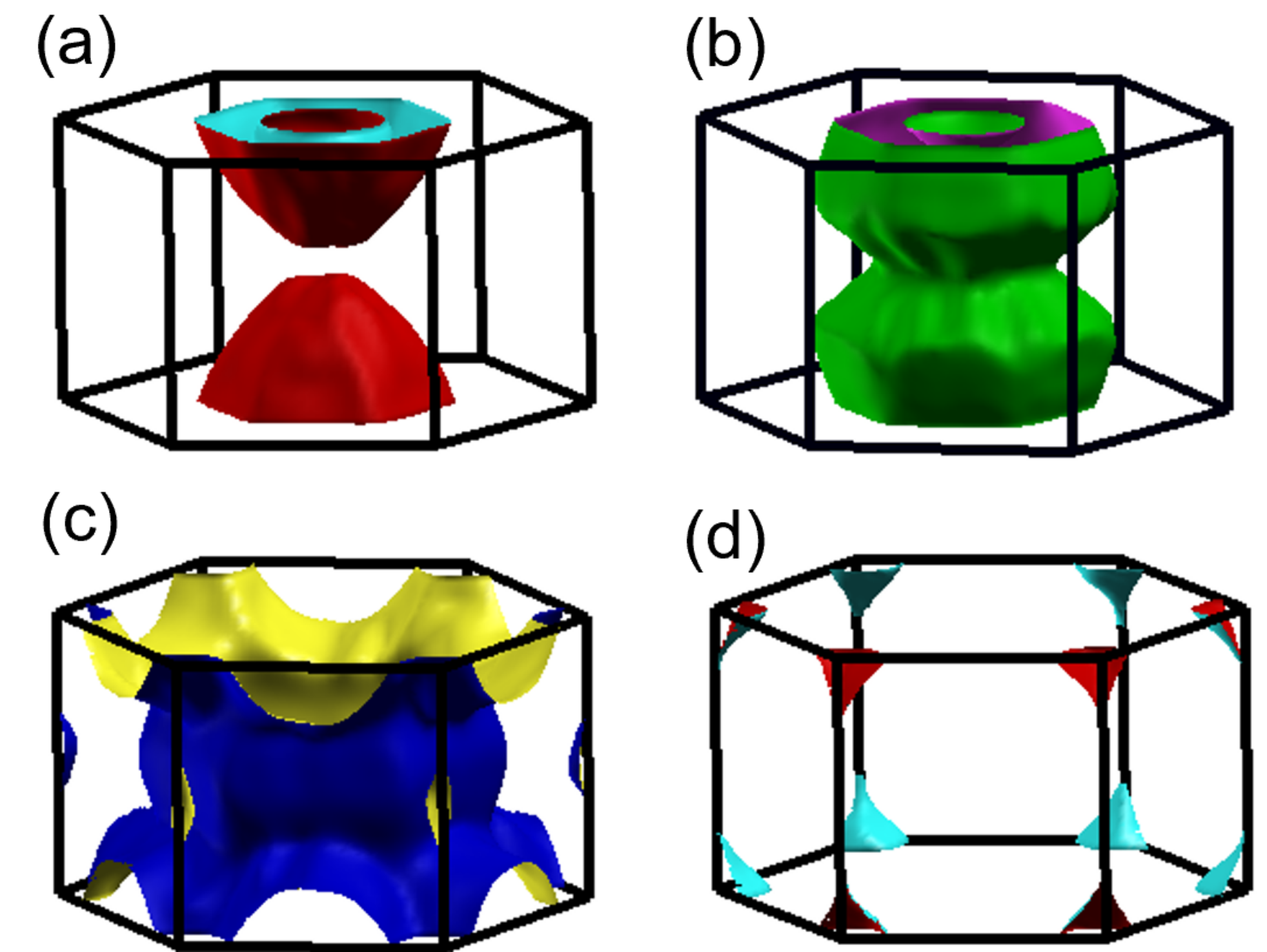}
\caption{\label{figs17} Fermi surfaces for pristine IrSb from DFT calculations. (a)-(d) represent four bands crossing the Fermi surface. }
\end{figure*}

\begin{figure*}[t]%
\centering
\includegraphics[width=11cm]{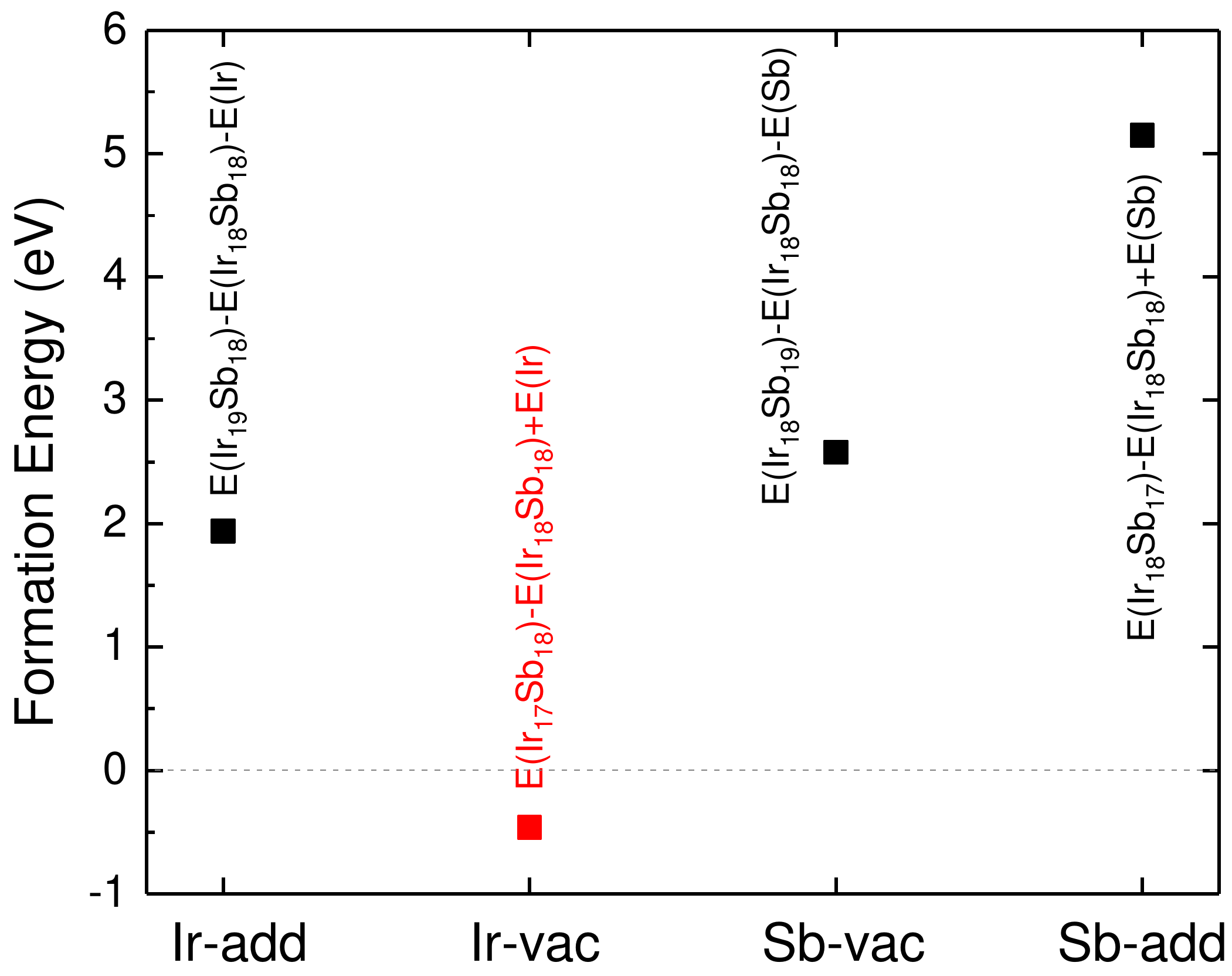}
\caption{\label{figs18} Formation energy of a vacancy or an interstitial atom of Ir or Sb. According to the (1/3, 1/3, 0) nesting vector of the Fermi surface, we first built a 3$\times$3$\times$1 supercell. Next, we considered one vacancy or one interstitial atom per 3$\times$3$\times$1 superstructure, where Ir-add, Ir-vac, Sb-vac, and Sb-add represent an Ir interstitial atom, an Ir vacancy, a Sb vacancy, and a Sb interstitial atom, respectively. As shown in the figure, only the formation energy of Ir-vac is below zero and is apparently smaller than those in the other situations, which means that the superstructure must stem from an ordered arrangement of an Ir vacancy(ies).}
\end{figure*}

\begin{figure*}[t]%
\centering
\includegraphics[width=11cm]{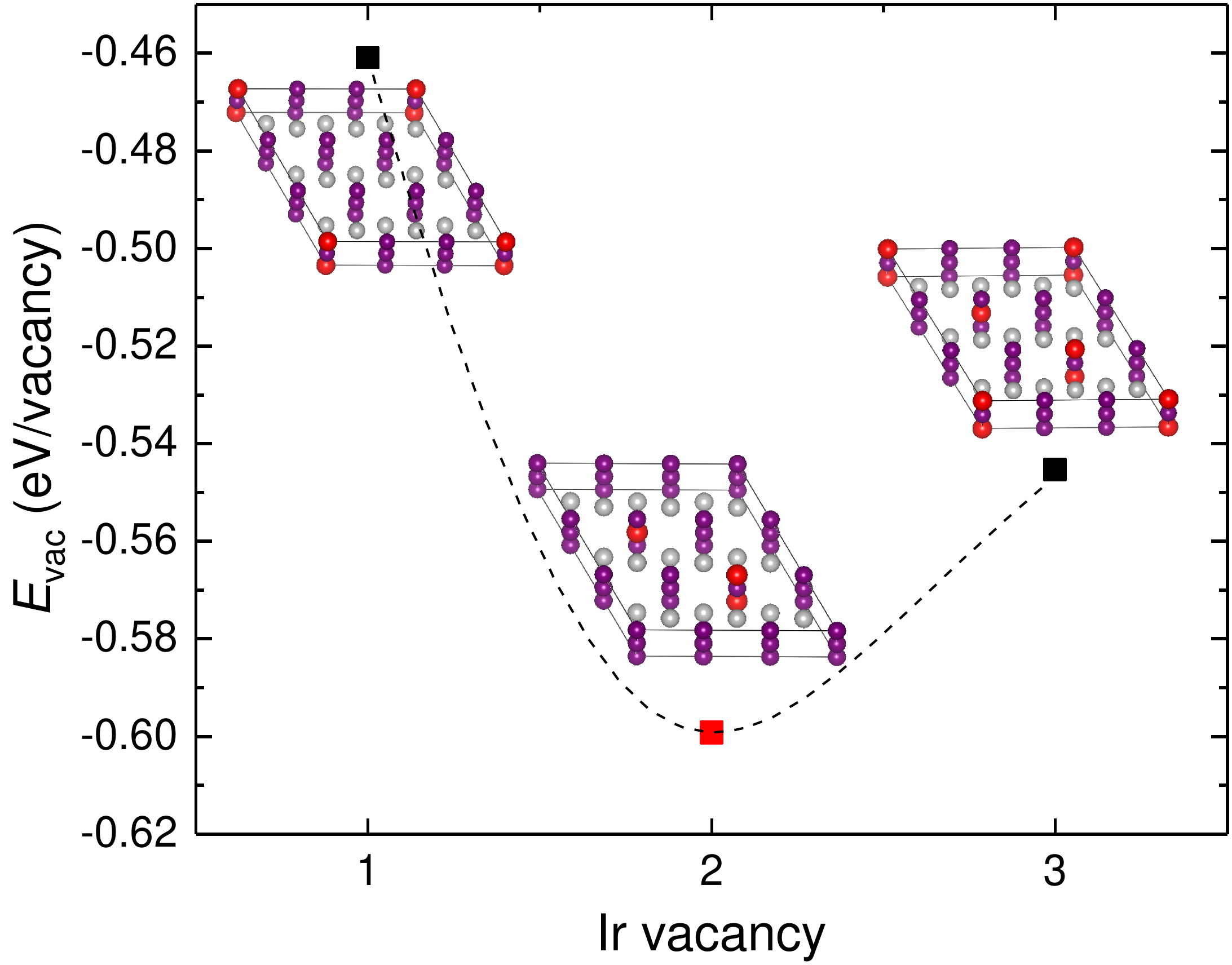}
\caption{\label{figs19}Vacancy formation energy vs number of Ir vacancies. The red, purple, and grey circles represent vacancies, iridium, and antimony, respectively. While there is only one configuration for the one vacancy case, several other possibilities arise in the case of three vacancies. The example shown in the figure possesses the lowest formation energy; however, the energy is still larger than that of the BHV ordering. An intuitive explanation for this parabolic energy evolution is as follows: According to Fig. S18, the system tends to form Ir vacancies to lower the energy, while excessive charged vacancies tend to repel each other and increase the total energy. Thus, the optimal balance is two vacancies per supercell.}
\end{figure*}

\begin{figure*}[t]%
\centering
\includegraphics[width=11cm]{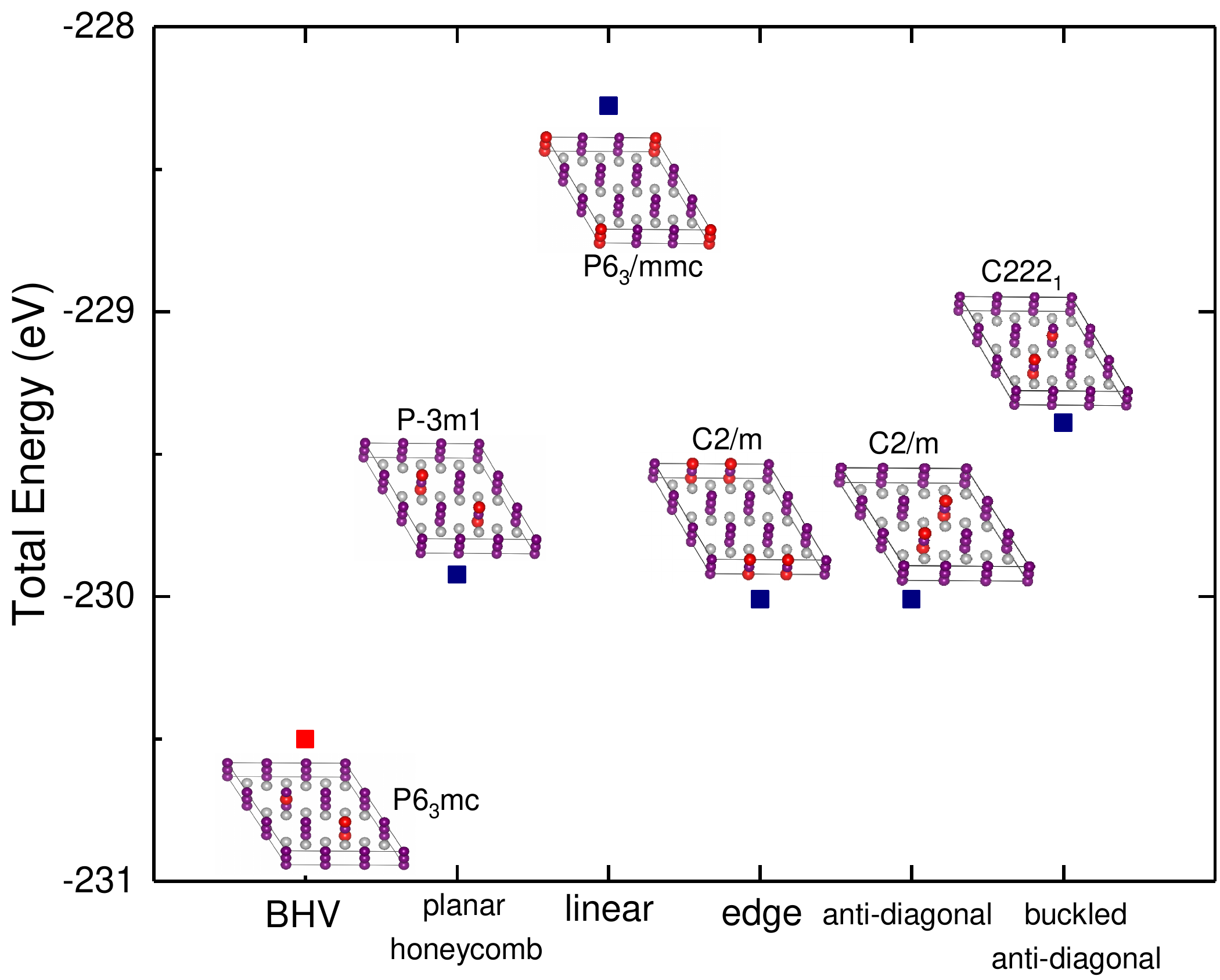}
\caption{\label{figs20}Total energy of different vacancy configurations. The red, purple, and grey circles represent vacancies, iridium, and antimony, respectively. In addition to the space groups allowed by the extinction rule of diffraction, other apparently highly symmetric space groups, such as the planar-honeycomb ($P$-$3m1$) and anti-diagonal configurations, are also included for comparison. Compared to the other configurations, the phase with BHV ordering has the lowest energy. Actually, for a given 3$\times$3$\times$1 NiAs-type superlattice, the structure with BHV ordering best maximizes the vacancy-vacancy distance, thereby minimizing the total energy.}
\end{figure*}

\begin{figure*}[t]%
\centering
\includegraphics[width=14cm]{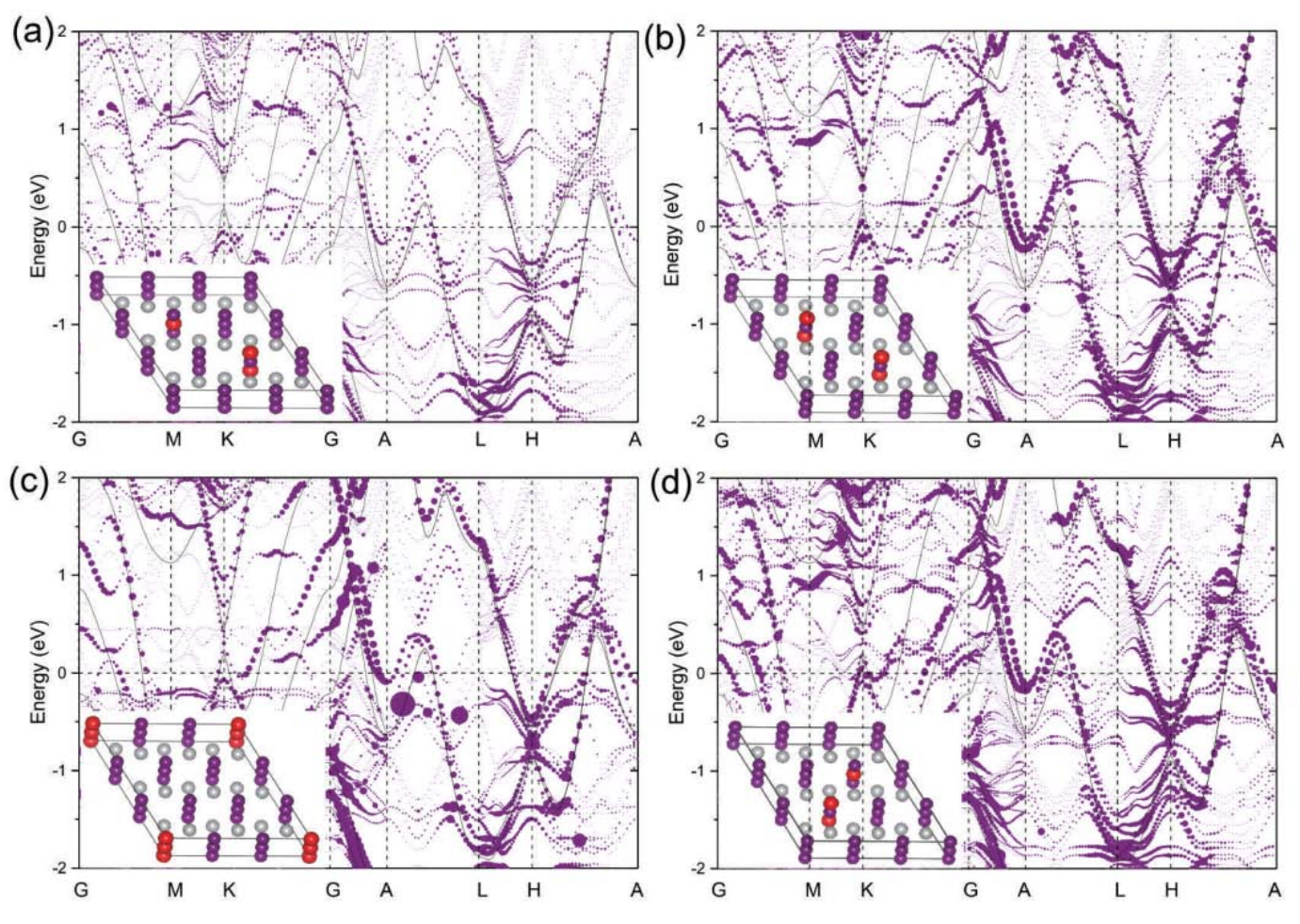}
\caption{\label{figs21} (a)-(d) Unfolded ordering structure in a pristine IrSb cell. The respective structures are illustrated in the insets. The purple dots denote the unfolded bands, while the band structure of the primary NiAs unit cell is superimposed as the black curves. The sizes of the circles are proportional to their spectral weights. Compared to the other configurations, the BHV ordering can mostly smear the bands.}
\end{figure*}

\bibliographystyle{ying}

\clearpage
\end{document}